\documentclass[superscriptaddress,nofootinbib,preprint,tightenlines]{revtex4-1}
\usepackage{epsfig}
\usepackage{graphicx}
\usepackage{psfrag}
\usepackage{amsmath,amssymb}
\usepackage{colordvi}
\usepackage{amsfonts}
\usepackage{enumerate}
\usepackage{slashed}
\usepackage{gensymb}
\usepackage{multirow}
\usepackage{color}

\newcommand{\p}{\partial}
\newcommand{\m}{\mathcal}
\newcommand{\ma}{\mathbf}
\newcommand{\e}{\epsilon}
\newcommand{\la}{\langle}
\newcommand{\ra}{\rangle}

\newcommand{\sla}{\slashed}

\newcommand{\boldgreek}[1]{{\mbox{\boldmath{$#1$}}}}
\newcommand{\bg}{\boldgreek}

\graphicspath{{./Diagrams/}}

\begin{document}
\title{Effective Angular Momentum Operators in NRQED \\ and Matching at One-Loop Order}
\author{Panying Chen}
\affiliation{Maryland Center for Fundamental Physics, Department of Physics, University of Maryland, College Park, MD 20742, USA}
\author{Xiangdong Ji}
\affiliation{Maryland Center for Fundamental Physics, Department of Physics, University of Maryland, College Park, MD 20742, USA}
\affiliation{INPAC, Department of Physics, Shanghai Jiao Tong University, Shanghai 200240, China}
\affiliation{Center for High-Energy Physics and Institute of Theoretical Physics, Peking University, Beijing 100871, China}
\author{Yue Zhang}
\affiliation{Abdus Salam International Centre for Theoretical Physics, Trieste 34100, Italy}

\date{\today}

\vspace{0.5in}
\begin{abstract}
We derive the effective angular momentum operator to $1/m^2$ and one-loop
order in non-relativistic quantum electrodynamics (NRQED). In both dimensional and
three-momentum-cutoff regularization schemes, we obtain the non-relativistic
expansion for the spin and orbital angular momentum operators of the
electron and photon, respectively. Our result is useful in
precision calculations of the angular momentum properties for
non-relativistic QED bound states, such as atom systems.

\end{abstract}
\preprint{\vbox{\hbox{IC/2010/001}}}

\maketitle



\section{Introduction}

The spin structure of bound states is a topic of considerable
interest in recent years \cite{Filippone:2001ux}.
The so-called ``proton spin crisis'' was generated from the EMC
experiment~\cite{EMC} in which it was found only a small fraction of the total
proton spin is carried by the quark spin. In Ref.~\cite{Ji:1996ek}, a
gauge-invariant decomposition of the nucleon angular momentum into
the quark spin, orbital and the gluon contributions has been
introduced. The connection between off-forward (or generalized) parton distributions and the
quark/gluon angular momenta within the proton has been established. The
total angular momentum carried by quarks is measurable
through the deeply-virtual Compton scattering (DVCS)~\cite{Ji:1996nm}. A large amount
of works have also been done in studying the angular momentum carried
by the gluons, $\Delta G$, and several experiments have been
designed to measure this quantity and interesting physics
has been learnt~\cite{data}.

While much of the attention has been focused on
relativistic theory, it is illuminating to study
angular momentum in the non-relativistic limit of QED/QCD.
A clear understanding of the non-relativistic angular momentum structure is
useful for many systems, including the hydrogen-like
atoms~\cite{Eides:2000xc} and the heavy
quarkonium~\cite{Hoang:2002ae} bound states. The fraction of angular momentum
carried by the photon in hydrogen atom is relevant to the gluon contribution to
the spin of the proton. Since the hydrogen atom is a
non-relativistic bound state, the naive perturbation theory of
relativistic QED is not applicable because multiple energy scales
present. For the ground state, the
electron has orbital velocity $v \sim \alpha_{\rm _{EM}}$ where
$\alpha_{\rm _{EM}}\approx 1/137$ is the electromagnetic
fine-structure constant, therefore the loop expansion in
$\alpha_{\rm _{EM}}/v$ does not converge~\cite{weinberg}. As
another example, in production of electron-positron pairs near the threshold,
the interactions are always accompanied with the famous Sommerfeld
enhancement proportional to $\alpha_{\rm _{EM}}/v$. Attempts to calculate
these corrections within the relativistic theory is difficult analytically,
usually involving numerical solution of the Bethe-Salpeter equations.

Effective field theory approach sheds new light on the problems described
above. In recent years, effective theories
including non-relativistic QED (NRQED)~\cite{Caswell:1985ui} and the non-relativistic QCD (NRQCD), has
become a usual tool in solving non-relativistic bound
states.
NRQED has been used to calculate the hyperfine splitting and lamb
shift of the hydrogen system with considerable simplification~\cite{Pineda:1997ie, Kinoshita:1995mt, Luke:1999kz}.
NRQCD has
been used in analyzing the heavy quarkonium inclusive production in
colliders~\cite{Beneke:1997zp} and precision bound-state
calculations on the lattice. By observing that the Coulomb
interactions exchanging momentum at much lower scale compared to
the fermion mass $m$, one can integrate out the large momentum scale,
resulting in higher dimensional effective operators suppressed by
powers of ${p}/{m}\ll1$. In effective theories, relativistic
and radiative corrections to bound state problems can be
systematically expanded in terms of ${p}/{m}$ and $\alpha_{\rm
_{EM}}$. The non-relativistic bound state wavefunction properly sums
up all order corrections due to the Coulomb photon exchange.
Therefore, the NRQED as an effective field theory, is capable
of describing the hydrogen atom up to any definite order with desired
precision in a language familiar in quantum mechanics.

In the non-relativistic limit, the orbital angular momentum $\vec{L}$ and
spin $\vec{S}$ of the electron are both conserved and can be
used to classify the energy eigenstates. Moreover, the contribution from the
gauge potential in the orbital part is proportional to the velocity
and hence negligible. $\vec{L}$ is then the usual non-relativistic
angular momentum. When taking into account the relativistic effects and
quantum corrections, the angular momentum operator becomes more complicated.
For example, for relativistic Dirac Hamiltonian, neither $L$
nor $S$ maintains as a good quantum due to corrections of order $v^2/c^2$.
These are relativistic corrections starting at order $\m{O}(\alpha_{\rm _{EM}}^2)$.
Radiative corrections will further contribute at order $\m{O}(\alpha_{\rm
_{EM}}^3)$~\cite{Chen:2009rw}.

In this paper, we extend the discussions in~\cite{Chen:2009rw} and present relevant details of 
deriving the effective angular momentum operator in NRQED. 
For this purpose, we will construct a set of
gauge-invariant effective operators in the non-relativistic
theory. The matrix elements of the spin (orbital) angular momentum
for the electron and the gauge fields are calculated in full QED and
matched to the effective theory up to order ${\alpha_{\rm
_{EM}}}/{m^2}$. The main results of this paper are
Eqs.~(\ref{spin})--(\ref{gamma}) and Eqs.~(\ref{WilsonBilinear}),
(\ref{WilsonGauge}), (\ref{cutoff_Lag}). Using the power counting in
NRQED~\cite{Luke:1999kz, Bodwin:1994jh}, we have applied the
one-loop matching result to calculate the angular momentum carried by radiative
photons in the hydrogen atom up to $\mathcal{O}(\alpha_{\rm
_{EM}}^3)$ in Ref.~\cite{Chen:2009rw}. The further extension to the
angular momentum decomposition in QCD/NRQCD will be presented elsewhere.

The NRQED Lagrangian is first given in Ref.~\cite{Caswell:1985ui}.
In Ref.~\cite{Bodwin:1994jh}, a power counting rule has been
established.
\begin{eqnarray}
\label{NRQED Lagrangian}
 \m{L}_{\rm NRQED} & = & \psi^\dag\left\{iD^0 + c_2\frac{\mathbf{D}^2}{2m} + c_4\frac{\mathbf{D}^4}{8m^3} + \frac{e\,c_F}{2m}\bg{\sigma}\cdot\mathbf{B} + \frac{e\,c_D}{8m^2} [\bg{\nabla}\cdot\mathbf{E}]  + \frac{ie\,c_S}{8m^2}\bg{\sigma}\cdot(\mathbf{D}\times\mathbf{E}-\mathbf{E}\times\mathbf{D}) \right\}\psi \nonumber\\
 & & - \frac{d_1}{4} F_{\mu\nu}F^{\mu\nu} + \frac{d_2}{m^2} F_{\mu\nu}\square F^{\mu\nu} + \m{L}_{\rm 4-Fermi} + \cdots \ ,
\end{eqnarray}
here $\m{L}_{\rm 4-Fermi}$ represents all the four-fermi terms in
the Lagrangian and will not be used in this paper. The Wilson
coefficients in the effective Lagrangian are well
known~\cite{Manohar:1997qy}. Up to $\m{O}(\alpha_{\rm _{EM}})$
order,
\begin{eqnarray}
  &&c_2 = c_4 = 1;\;\; c_F = 1 + \frac{\alpha_{\rm _{EM}}}{2\pi};\;\; c_D = 1 - \frac{4\alpha_{\rm _{EM}}}{3\pi} \ln\frac{\mu^2}{m^2}   ;\;\; c_S = 1 + \frac{\alpha_{\rm _{EM}}}{\pi};\nonumber\\
  &&d_1 = 1 + \frac{\alpha_{\rm _{EM}}}{3\pi}\ln\frac{\mu^2}{m^2};\;\; d_2 = \frac{\alpha_{\rm _{EM}}}{60\pi}\ .
\end{eqnarray}
In the calculation, we will use the ``old-fashioned''
NRQED~\cite{holstein}, which is equivalent to the intermediate
effective theory in pNRQED~\cite{Brambilla:1999xf}. In performing
the matching, $1/m$ expansion of the matrix elements (so are the
effective operators) as described in Ref.~\cite{Manohar:1997qy} is
understood. It is proved in Ref.~\cite{Grinstein:1997gv} that this
agrees with the multi-pole expansion method in
Ref.~\cite{Labelle:1996en}.  More explanations will follow along
with the detailed calculations.

\section{Angular Momentum Operators in QED and NRQED and Matching Conditions}


The question we are interested in is how the total angular momentum is distributed among
various components, e.g., that carried by electron and photon, respectively.
The effective non-relativistic angular momentum operators are useful
in studying many non-relativistic bound state systems, including the
hydrogen-like atoms and heavy quarkonia.  The purpose of the following
sections will be the construction of gauge-invariant angular
momentum operators in NRQED. We start by first reviewing the angular momentum
operator in full QED. We then use the Foldy-Wouthysen (FW) transformation to get the
corresponding NRQED operators at tree level. We then use the N\"{o}ether current
method to derive the total NRQED angular momentum operator from the effective
NRQED lagrangian. This method is accurate to all loops. However,
it does not give us the results for individual relativistic angular momentum
components. Therefore, we write down the Wilson expansion
of these angular momentum components in terms of the non-relativistic
angular momentum operators. In the following sections,
we determine the Wilson coefficients using the matching method.

\subsection{Angular Momentum in Full QED}

It is straightforward to derive the total angular momentum operator
from the QED Lagrangian using the N\"{o}ether current method.
Starting from
\begin{equation}
  \m{L}_{\rm QED} = -\frac{1}{4} F^{\mu\nu}F_{\mu\nu} + \bar{\Psi}(i\sla{D} - m)\Psi \ ,
\end{equation}
the conserved QED angular momentum is~\cite{Jaffe:1989jz}
\begin{eqnarray}
\label{QED_J}
\mathbf{J}_{\rm QED} & = & \int d^3x \left\{\Psi^\dag \frac{\bg{\Sigma}}{2}\Psi + \Psi^\dag (\mathbf{x} \times \bg{\pi})\Psi + \mathbf{x}\times (\mathbf{E}\times \mathbf{B})\right\}  \equiv  \mathbf{S}_q+\mathbf{L}_q+\mathbf{J}_\gamma \ ,
\end{eqnarray}
in which
\begin{eqnarray}
\label{QED_Op}
\mathbf{S}_q & \equiv & \frac{1}{2}\int d^3x \Psi^\dag \bg{\Sigma} \Psi \ , \nonumber\\
\mathbf{L}_q & \equiv & \int d^3x \Psi^\dag (\mathbf{x} \times \bg{\pi}) \Psi \ , \nonumber\\
\mathbf{J}_\gamma & \equiv & \int d^3x \;\mathbf{x}\times (\mathbf{E}\times \mathbf{B}) \ .
\end{eqnarray}
The individual relativistic operators $\mathbf{S}_q, \mathbf{L}_q,
\mathbf{J}_\gamma$ are gauge invariant and are regarded as the
electron spin, electron orbital angular momentum and photon angular
momentum, respectively. Alternative separation can be achieved
in some particular gauge and frame choices.

Our goal is to construct the non-relativistic counterparts
to $\mathbf{S}_q, \mathbf{L}_q,$ and $\mathbf{J}_\gamma$.
The result within NRQED is gauge
invariant, with the manifest separation of high- and
low-scale physics.

\subsection{Non-Relativistic Reduction: Foldy-Wouthysen Transformation}


In this subsection, we apply the Foldy-Wouthysen (FW) transformation
to the QED operators in Eq.~(\ref{QED_Op}). We shall keep
in mind that the FW transformation does not yield any information
about radiative corrections, due to its quantum mechanical nature.
Consequently, one can obtain only the leading-order Wilson
coefficients by transforming the relativistic operators
$\mathbf{S}_q$, $\mathbf{L}_q$, and $\mathbf{J}_\gamma$, respectively.
Any Wilson coefficient starting from order $\m{O}(\alpha_{\rm
_{EM}})$ or higher would not show up in the transformation.
Furthermore, the FW transformation only rotate the electron
wavefunction, leaving the photon sector $\ma{J}_\gamma$ unchanged.

Here is a quick review of the FW transformation. The Dirac
Hamiltonian describes the relativistic fermion containing both
positive and negative frequencies. At lower energy, the negative
frequency part (positron) is subdominant.
For an electron state,  the physics is dominated by the interactions
of the electron itself, and the correction from the positron is
suppressed by the mass. The latter is reflected by the
off-block-diagonal elements in the Hamiltonian. The FW
transformation utilizes a unitary transformation to
block-diagonalize the Dirac Hamiltonian up to order $\m{O}(m^{-2})$.
The unitary transformation matrix is
\begin{equation}
\label{FoldyU}
U = \exp\left[\frac{\beta \bg{\alpha} \cdot \bg{\pi}}{2m}\right] = \exp\left[\frac{\bg{\gamma} \cdot \bg{\pi}}{2m}\right] \ ,
\end{equation}
where in the Dirac representation
\begin{equation}
  \beta  = \left[\begin{array}{cc} 1 & 0 \\ 0 & -1 \end{array}\right], \, \, \, \, \,
  \bg{\alpha} = \left[\begin{array}{cc} 0 & \bg{\sigma} \\ \bg{\sigma} & 0 \end{array}\right] \ .
\end{equation}
One can use $U$ to eliminate the lower components in a
four-component Dirac spinor of electron up to order $\m{O}(m^{-3})$ and a
quantum mechanics operator $\hat{\mathcal{O}}$ transforms
accordingly
\begin{eqnarray}
\label{FoldyO}
  u(\mathbf{p}) & \to & U \cdot u(\mathbf{p}) = {u_h(\mathbf{p}) \choose 0} + \m{O}(m^{-3}) \ ,\nonumber\\
  \hat{\m{O}} & \to & U \cdot \hat{\m{O}} \cdot U^\dag \ .
\end{eqnarray}
The operators in Eq.~(\ref{QED_Op}) take the following form under
the FW transformation
\begin{eqnarray}
\label{FW}
\int d^3 \mathbf{x}\ \Psi^\dag \frac{\bg{\Sigma}}{2} \Psi
& \rightarrow & \int d^3 \mathbf{x} \psi^\dag\left\{
\frac{\bg{\sigma}}{2} + \frac{1}{8m^2} \left[
(\bg{\sigma} \times \bg{\pi})\times \bg{\pi} - \bg{\pi} \times
(\bg{\sigma} \times \bg{\pi}) \right]
+ \frac{e\mathbf{B}}{4m^2} \right\}\psi \ , \nonumber \\
\int d^3 \mathbf{x}\ \Psi^\dag \mathbf{x} \times \bg{\pi} \Psi
& \rightarrow & \int d^3 \mathbf{x} \psi^\dag\left\{
\mathbf{x} \times \bg{\pi} - \frac{1}{8m^2}  \left[
(\bg{\sigma} \times \bg{\pi})\times \bg{\pi} - \bg{\pi} \times
(\bg{\sigma} \times \bg{\pi}) \right]
 - \frac{e\mathbf{B}}{4m^2} \right\}\psi \nonumber \\
& & + \int d^3 \mathbf{x}
\frac{e}{8m^2} \left\{ \mathbf{x} \times [\mathbf{B} \times
\bg{\nabla}(\psi^\dag \psi)] + \mathbf{x} \times  \psi^\dag \left[\mathbf{B} \times(\bg{\sigma} \times \overleftrightarrow{\bg{\pi}}) \right]\psi  \right\} \ , \nonumber \\
\int d^3x \;\mathbf{x}\times (\mathbf{E}\times \mathbf{B})  & \rightarrow & \int d^3x \;\mathbf{x}\times (\mathbf{E}\times \mathbf{B}) \ ,
\end{eqnarray}
where $\psi$ is the non-relativistic electron field and
$\overleftrightarrow{\bg{\pi}}$ is defined as:
\begin{eqnarray}
\psi^\dag_{p'}\overleftrightarrow{\bg{\pi}}\psi_{p} & \equiv & \psi^\dag_{p'} (\bg{\pi}\psi_{p}) + \big(\bg{\pi}\psi_{p'}\big)^\dag \psi_{p} = \psi^\dag_{p'} \left( \mathbf{p} + \mathbf{p}' -2e\mathbf{A}\right) \psi_{p} \ .
\end{eqnarray}

In the above results, the next-to-leading order relativistic
corrections has been properly taken into account. 
In Ref.~\cite{Chen:2009rw}, it has been shown that the relativistic correction to the orbital angular momentum 
of ground state hydrogen atom can be calculated with the above leading order operators. 
As we will see in the following section, these results are also consistent with those
derived from matching between QED and NRQED at $\mathcal{O}(\alpha_{_{\rm EM}}/m^2)$ order.


\subsection{Total Angular Momentum from NRQED Lagrangian}

In this subsection, we will first derive $\ma{J}_{\rm NRQED}$ using
the N\"{o}ether current method. The NRQED Lagrangian contains a
second-order-derivative term $\frac{d_2}{m^2} F_{\mu\nu}\square
F^{\mu\nu}$ and calls for special treatments. In Appendix A, we
present the generalized formalism to derive the equation of motion
and the conserved currents, including the energy-momentum current
$T^{\mu\nu}$ and the angular momentum current
$\mathcal{M}^{\mu\nu\lambda}$, from a general Lagrangian
$\m{L}=\m{L}(x, \phi, \partial_\mu\phi,
\partial_\mu\partial_\nu\phi)$. For NRQED, the equation of motion
with regard to $A^0$ field is
\begin{eqnarray}
-e\psi^\dagger\psi - \frac{e\,c_D}{8m^2}\bg{\nabla}^2(\psi^\dag
\psi) + \frac{ie\,c_S}{4m^2}\bg{\nabla} \cdot \left[\psi^\dag
\left(\bg{\sigma}\times\mathbf{D}\right)\psi\right] + d_1 \p_\mu
F^{\mu 0} - \frac{4d_2}{m^2}\square\left(\p_\mu F^{\mu 0}\right) = 0
. \label{EoM}
\end{eqnarray}
and the total angular momentum
\begin{eqnarray}
\label{NRQED_J} \mathbf{J}_{\rm NRQED} & = & \int d^3 \mathbf{x}
\left\{ \psi^\dag \left(\frac{\bg{\sigma}}{2}\right) \psi +
\psi^\dag \left(\mathbf{x} \times \bg{\pi}\right) \psi +
\frac{e\,c_D}{8m^2} \mathbf{x} \times [\mathbf{B} \times
\bg{\nabla}(\psi^\dag \psi)] \rule{0cm}{6mm}\right.
\nonumber\\
 &+&\frac{e\,c_S}{8m^2} \mathbf{x} \times \psi^\dag \left[\mathbf{B}
\times(\bg{\sigma} \times \overleftrightarrow{\bg{\pi}}) \right]\psi  + d_1\,\mathbf{x} \times
\left( \mathbf{E} \times \mathbf{B} \right) \nonumber\\
 &+&\left.\left(-\frac{4d_2}{m^2}\right)\left[ - \mathbf{x} \times
( \square \mathbf{E} \times \mathbf{B} ) + \dot{\mathbf{E}}^a (\mathbf{x}\times \bg{\nabla})
 \mathbf{E}^a - \dot{\mathbf{B}}^a(\mathbf{x}\times \bg{\nabla}) \mathbf{B}^a  +
 \dot{\mathbf{E}}\times\mathbf{E} - \dot{\mathbf{B}}\times\mathbf{B}\right] \right\} \nonumber\\
 &+& \cdots \ .
\end{eqnarray}
We will not consider four-fermion operators in this paper.
The total angular momentum operator $\mathbf{J}_{\rm NRQED}$ in NRQED is manifestly gauge invariant.
However, it is difficult to see a clear correspondence with the individual relativistic operators.
Therefore, we shall use the matching method to derive the non-relativistic expansion
of the relativistic operators.

\subsection{Relativistic Angular Momentum Operator in NRQED}

We observe that the effective Lagrangian in Eq.~(\ref{NRQED
Lagrangian}) still possesses the rotational symmetry in 3D,
therefore implies the existence of conserved total angular momentum
$\ma{J}_{\rm NRQED}$~\footnote[1]{The authors in
Ref.~\cite{Brambilla:2003nt} have noticed this as well, and derived
the angular momentum in the NRQED up to order
$\m{O}\left(\alpha_{\rm _{EM}}^0/m^0\right)$.}. In this paper, we
will derive non-relativistic effective angular momentum
operators up to order $\m{O}\left(\alpha_{\rm
_{EM}}/m^2\right)$, through order-by-order matching to the full
QED.

The general form of angular momentum operators in
NRQED can be constructed as follows. First, we write down all the
axial-vector Hermitian operators respecting gauge and 3D rotational
invariance. Second, we require these operators contain no time
derivatives on the fermion fields. The removal of $D^0$ can be done
by field redefinition as shown in~\cite{Manohar:1997qy}. Any
operator containing $D^0$ can be rewritten as linear combinations of
other operators by the equation of motion without changing the
on-shell matrix elements. Third, there will be non-local operators
in $\ma{L}_q$ and $\ma{J}_\gamma$, containing the space coordinate
$\ma x$ explicitly.

Here we write down the general form of the effective operators, with
the Wilson coefficients to be determined. For electron spin,
\begin{eqnarray}\label{spin}
\mathbf{S}_q \rightarrow \mathbf{S}_q^{\rm eff} & = &
\int d^3x\left\{\rule{0cm}{7mm}a_\sigma\psi^\dag\frac{\bg{\sigma}}{2}\psi
+ \frac{a_\pi}{8m^2} \psi^\dag \left[(\bg{\sigma}\times\bg{\pi}) \times
\bg{\pi}-\bg{\pi} \times (\bg{\sigma}\times\bg{\pi})\right]\psi +
\frac{e\,a_B}{4m^2}\psi^\dag\mathbf{B}\psi \rule{0cm}{7mm}\right\} \nonumber\\
 & & +\int d^3x \left\{\frac{a_{\gamma_1}}{m^2}\left[\dot{\mathbf{E}}\times\mathbf{E}
- \dot{\mathbf{B}}\times\mathbf{B}\right] + \frac{a_{\gamma_2}}{m^2}
\left[(\bg{\nabla}\cdot \ma{E})\ma{B} - (\ma{B}\cdot \bg{\nabla}) \ma{E} \rule{0cm}{4mm}\right]
\rule{0cm}{7mm}\right\} + \m{O}(m^{-3}) \ , \nonumber \\
\end{eqnarray}
for electron orbital momentum,
\begin{eqnarray}\label{orbit}
\mathbf{L}_q \rightarrow \mathbf{L}_q^{\rm eff} & = & \int d^3x\left\{\rule{0cm}{7mm} d_\sigma\psi^\dag\frac{\bg{\sigma}}{2}\psi + \frac{d_\pi}{8m^2} \psi^\dag \left[(\bg{\sigma}\times\bg{\pi}) \times \bg{\pi}-\bg{\pi} \times (\bg{\sigma}\times\bg{\pi})\right]\psi \right.\nonumber\\
 & & + d_R \psi^\dag(\mathbf{x}\times \bg{\pi})\psi + \frac{e\,d_B}{4m^2}\psi^\dag\mathbf{B}\psi + \frac{e\,d_D}{8m^2}  \mathbf{x}\times \left[\mathbf{B}\times \bg{\nabla}(\psi^\dag\psi)\right]\nonumber\\
 & & + \frac{e\,d_S}{8m^2}{\mathbf x}\times \psi^\dag \left[\mathbf{B} \times(\bg{\sigma} \times \overleftrightarrow{\bg{\pi}}) \right]\psi + \frac{e\,d'_S}{8m^2}{\mathbf x}\times \psi^\dag \left[\bg{\sigma}\times (\mathbf{B} \times
\overleftrightarrow{\bg{\pi}}) \right]\psi\nonumber\\
 & & \left.+ \frac{e\,d_E}{4m}\mathbf{x} \times\psi^\dag\left(\bg{\sigma} \times \mathbf {E}\right)\psi \rule{0cm}{7mm}\right\} + \int d^3x \left\{\rule{0cm}{7mm}d_\gamma\, \mathbf{x}\times(\mathbf{E}\times\mathbf{B}) + \frac{d_{\gamma_1}}{m^2}\left[\dot{\mathbf{E}}\times\mathbf{E} - \dot{\mathbf{B}}\times\mathbf{B}\right]\right.\nonumber\\
 & &+ \frac{d_{\gamma_2}}{m^2}
\left[(\bg{\nabla}\cdot \ma{E})\ma{B} - (\ma{B}\cdot \bg{\nabla}) \ma{E} \rule{0cm}{4mm}\right] + \frac{d_{\gamma_3}}{m^2} \left[ - \mathbf{x} \times ( \square \mathbf{E} \times \mathbf{B} ) \rule{0cm}{4mm}\right] \nonumber\\
 & & \left.+ \frac{d_{\gamma_4}}{m^2}\left[\dot{\mathbf{E}}^a (\mathbf{x}\times \bg{\nabla}) \mathbf{E}^a - \dot{\mathbf{B}}^a (\mathbf{x}\times \bg{\nabla}) \mathbf{B}^a\right] \rule{0cm}{7mm}\right\} +\m{O}(m^{-3}) \ ,
\end{eqnarray}
and for photon angular momentum operator,
\begin{eqnarray}\label{gamma}
\mathbf{J}_\gamma \rightarrow \mathbf{J}_\gamma^{\rm eff} & = & \int d^3x\left\{\rule{0cm}{7mm}
f_\sigma\psi^\dag\frac{\bg{\sigma}}{2}\psi
+ \frac{f_\pi}{8m^2} \psi^\dag \left[(\bg{\sigma}\times\bg{\pi})
\times \bg{\pi}-\bg{\pi} \times (\bg{\sigma}\times\bg{\pi})\right]\psi \right.\nonumber\\
& & + f_R \psi^\dag({\mathbf x\times \bg{\pi}})\psi +
\frac{e\,f_B}{4m^2}\psi^\dag \mathbf{B} \psi
+ \frac{e\,f_D}{8m^2}  {\mathbf x}\times \left[\mathbf B \times \bg{\nabla}(\psi^\dag\psi)\right] \nonumber\\
& & + \frac{e\,f_S}{8m^2}{\mathbf x}\times \psi^\dag \left[\mathbf{B} \times(\bg{\sigma} \times \overleftrightarrow{\bg{\pi}}) \right]\psi + \frac{e\,f'_S}{8m^2}{\mathbf
x}\times \psi^\dag \left[\bg{\sigma}\times (\mathbf{B} \times
\overleftrightarrow{\bg{\pi}}) \right]\psi\nonumber\\
& & \left. +
\frac{e\,f_E}{4m}\mathbf{x} \times\psi^\dag\left(\bg{\sigma} \times \mathbf {E}\right)\psi \rule{0cm}{7mm}\right\} + \int d^3xf_\gamma\mathbf{x}\times(\mathbf{E}\times\mathbf{B})+ \m{O}(m^{-3}) \ .
\end{eqnarray}
In what follows,
we will perform the matching of various matrix elements between QED
and NRQED, to determine all the Wilson coefficients in
Eq.~(\ref{spin})--(\ref{gamma}).

The matching condition is
\begin{equation}
\label{MatchingCondition}
\la\mathbf{J}(\mu)\ra_{\rm QED} = \la\mathbf{J}^{\rm eff}(\mu)\ra_{\rm NRQED}\ .
\end{equation}
Although the primary application of our NRQED effective operators
involves bound states, we are not obliged to use a bound
state to do the matching.
For local operators which do not contain space coordinate $\ma{x}$ explicitly, we shall calculate two- and three-body matrix elements between plane-wave external states. On the other hand, when calculating the matrix element of non-local operators, such as $\ma{L}_q$, $\ma{J}_\gamma$ etc, we have to replace the simple plane waves by general wave packets.


An important remark is in order. The total angular momentum operator in Eq.~(\ref{NRQED_J}), originates from the underlying $SO(3)$ rotational symmetry of the NRQED Lagrangian. It has nothing to do with the perturbative expansion and must be exact. The gauge-invariant effective operators, as will be derived from the matching to the full theory, must agree with Eq.~(\ref{NRQED_J}) in the sum. 
This fact imposes the constraint $\ma{S}_q^{\rm eff} + \ma{L}_q^{\rm eff} + \ma{J}_\gamma^{\rm eff} = \ma{J}_{\rm NRQED}$, which in turn, leads to the sum rules among the above Wilson coefficients, as given explicitly in Ref.~\cite{Chen:2009rw}.


\section{Two-Body Matching Through Single-Electron States}

In this section, we consider matching the relativistic and NR angular momentum operators through single-electron
states. This step alone cannot determine all the
Wilson coefficients in Eqs.~(\ref{spin})--(\ref{gamma}). However, the calculation is
useful to illustrate the general procedure, and shows simple physical concepts behind
the matching. The result in this section is complementary to the three-body matching in
the next one.

We calculate the forward matrix elements of the angular momentum operators: $\la e_{p}| \m{O}_J|e_{p}\ra \equiv \la\m{O}_J\ra_{p,p}$,
with $\m{O}_J$ representing one of the relativistic operators $\ma{S}_q$, $\ma{L}_q$ and $\ma{J}_\gamma$ and $|e_p \ra$ is a free
electron state with momentum $p$.  We will perform
non-relativistic reduction of their matrix elements in QED, and match them with those in NRQED.

The spin operator $\ma{S}_q$ is local and its contribution can be easily calculated in QED. At tree level, we have
\begin{equation}
\la \ma{S}_q \ra_{p,p} = \frac{1}{2} \bar u(p) \bg{\Sigma} u(p) ,
\end{equation}
where $u(p)$ is the usual Dirac spinor.
The calculation of $\la\ma{L}_q\ra$ and $\la\ma{J}_\gamma\ra$ is more involved. The explicit presence of space coordinate $\ma{x}$ in the operator obscures the definition of the forward matrix element. Before going into explicit calculations, we address a basic question: what is the angular momentum for a free electron?

The idea of the total angular momentum of a plane wave is ill defined.
To see this, we first consider an off-forward matrix element of the operator $\ma{L}_q$. After some algebra,
\begin{eqnarray}
  \la \ma{L}_q \ra_{p',p} & = & \int d^3x \la \ma{p}'| \Psi^\dag (\ma{x}\times\bg{\pi} ) \Psi |\ma{p}\ra\nonumber\\
  & \equiv & \int d^3x e^{-i\ma{x}\cdot(\ma{p'}-\ma{p})}\ma{x} \times \ma{f}(p', p)\nonumber\\
  & = & (2\pi)^3[i\bg{\nabla}_{p'}\delta^3(\ma{p}-\ma{p}')] \times \ma{f}(p', p) \ ,
\end{eqnarray}
where $\ma{f}(p',p)=\bar u(p') \ma{p} u(p)$ at tree level.
The derivative on the delta function is singular if one takes $\ma{p}=\ma{p}'$.

A more realistic question would be to calculate the angular momentum of a distribution, or a {\it wave packet} $|\Phi \ra \equiv \int \frac{d^3 \ma{p'}}{(2\pi)^3} \Phi(\ma{p'})|\ma{p'}\ra$. The result will not only depend on the intrinsic properties of the electron itself, but also on
the shape of the wave packet. Such relevance arises because the angular momentum operator is a non-local operator.
In the momentum space, $\ma{x}$ can be Fourier transformed into the derivative over the conjugate momentum $\bg{\nabla}_{p'}$. After integration by parts, the contribution to the total angular momentum is therefore separated into the two parts: the derivative on the matrix element $\ma{f}(p',p)$ and on the wave packet $\Phi(\ma{p'})$,
\begin{eqnarray}\label{23}
  \la \ma{L}_q \ra_{\Phi} & = & \int d^3p \Phi^*(\ma{p}) \Phi(\ma{p})\left[i\bg{\nabla}_{p'} \times \ma{f}(p', p)\rule{0mm}{4mm}\right]_{p'=p} + \frac{1}{2}\int d^3p \bg{\nabla}_p\left[\Phi^*(\ma{p}) \Phi(\ma{p})\rule{0mm}{4mm}\right]\times \ma{f}(p,p).
\end{eqnarray}
In the first term, $[i\bg{\nabla}_{p'} \times \ma{f}(p', p)]_{p'=p}$ is the orbital angular momentum density intrinsic to the electron and contributes in the same manner as $\la\ma{S}_q\ra$.  The second term is precisely the angular momentum carried by the wave packet. Similar terms exist for the matrix elements of $\ma{J}_\gamma$, as well as any non-local effective operator. In the following, we will denote the class of contribution like the first term in Eq.~(\ref{23}) as the {\it local} contribution, and the second term as the {\it non-local} contribution for a matrix element between general wave packets. For local operators like $\ma{S}_q$, only the local contribution is present. For non-local operators, such as $\ma{L}_q$ and $\ma{J}_\gamma$ etc, both local and non-local parts of the matrix elements from the full theory and the effective theory should match, independently. This will provide us with enough information to obtain the relevant Wilson coefficients.

For non-relativistic reduction, we use the convention for a four-component spinor
\begin{equation}\label{nr}
u(p) = \sqrt{\frac{p^0+m}{2p^0}} {u_h \choose \frac{\mathbf{p} \cdot{\bg{\sigma}}}{p^0+m}u_h} \ ,
\end{equation}
where $u_h = \sqrt{2p^0}{1\choose0}$ or $\sqrt{2p^0}{0\choose1}$ and $p^0=\sqrt{\mathbf{p}^2+m^2}$.
So we can rewrite the full theory matrix elements in terms of $u_h$ and compare them with the effective theory. At tree level, the non-relativistic reduction of the QED matrix elements are
\begin{eqnarray}
\label{2bodytree}
&&\la \ma{S}_q \ra^{(0)}_{\rm local} = u_h^\dag
\frac{\bg{\sigma}}{2} u_h + \frac{1}{4 m^2} u_h^\dag \left[
(\bg{\sigma} \times \ma{p}) \times \ma{p} \right] u_h + \m{O}(m^{-3}) \nonumber \\
&&\la \ma{L}_q \ra^{(0)}_{\rm local} = - \frac{1}{4 m^2} u_h^\dag \left[
(\bg{\sigma} \times \ma{p}) \times \ma{p} \right] u_h + \m{O}(m^{-3}) \nonumber\\
&&\ma{f}(p,p)^{(0)}_{L_q,\rm non-local} = u_h^\dag\ma{p} u_h,
\end{eqnarray}
This simple result can be used to examine the FW-tranformed $\ma{S}_q$ and $\ma{L}_q$. By calculating the matrix elements of operators in Eq.~(\ref{FW}) between two non-relativistic spinor wave packets, we find they indeed agree with Eq.~(\ref{2bodytree}). Therefore, we conclude
\begin{equation}
  a_\sigma^{(0)} = a_\pi^{(0)} = -d_\pi^{(0)} = d_R^{(0)} = 1 \ ,
\end{equation}
at the leading order.

\begin{figure}[hbt]
\begin{center}
\includegraphics[width=4cm]{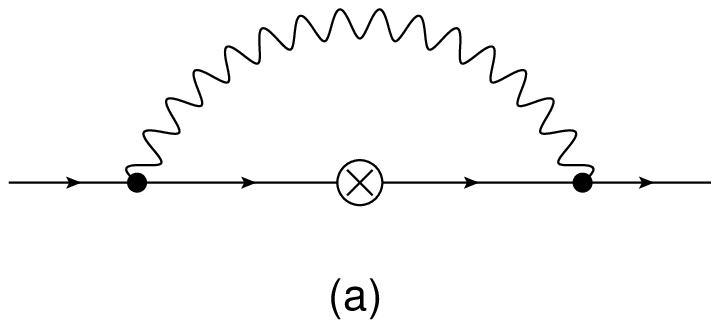}
\hspace{.5cm}
\includegraphics[width=4cm]{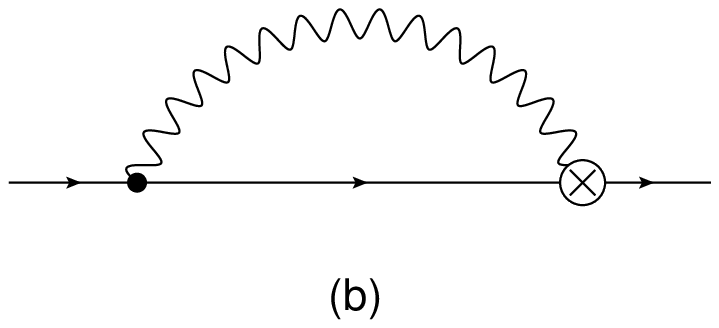}
\hspace{.5cm}
\includegraphics[width=4cm]{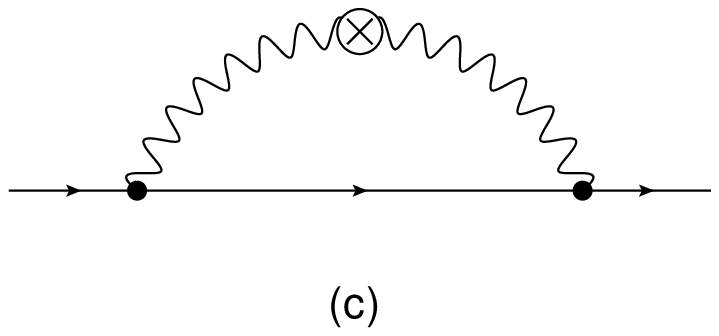}
\caption{One-loop corrections to $\la e|\mathbf{J}|e\ra$ in QED.  Here $\otimes$ can be either $\ma{L}_q$,$\ma{S}_q$, or $\ma{J}_\gamma$: (a) for $\ma{S}_q$; (a)(b) for $\ma{L}_q$; (c) for $\ma{J}_\gamma$.  Wave function renormalization diagrams, mass counterterms and the mirror diagrams are not shown explicitly. }
\label{QED_2body}
\end{center}
\end{figure}

The one-loop Feynman diagrams for the electron two-body matrix element of $\ma{S}_q$, $\ma{L}_q$ and $\ma{J}_\gamma$ are shown in Fig. \ref{QED_2body}. We use the dimensional regularization (DR) for both infrared (IR) and ultraviolet (UV) divergencies. The {\it local} contributions in the matrix elements are
\begin{eqnarray}
\label{2body1loopSpin}
  \la \ma{S}_q \ra^{(1)}_{\rm local} & = & \frac{\alpha_{\rm _{EM}}}{2 \pi} u_h^\dag\left\{ \frac{\bg{\sigma}}{2} + \frac{1}{4 m^2}  (\bg{\sigma} \times \ma{p}) \times \ma{p} \right\} u_h \nonumber\\
  \la \ma{L}_q \ra^{(1)}_{\rm local} & = & \frac{\alpha_{\rm _{EM}}}{2 \pi} u_h^\dag \left\{ \left( -\frac{4}{3\epsilon_{\rm UV}} - \frac{4}{3}\ln\frac{\mu^2}{m^2} - \frac{20}{9} \right) \frac{\bg{\sigma}}{2} -  \frac{1}{4 m^2} \frac{5}{3} (\bg{\sigma} \times \ma{p}) \times \ma{p} \right\}u_h \nonumber\\
  \la \ma{J}_\gamma \ra^{(1)}_{\rm local} & = & \frac{\alpha_{\rm _{EM}}}{2 \pi} u_h^\dag \left\{ \left( \frac{4}{3 \epsilon_{\rm UV}} + \frac{4}{3}\ln\frac{\mu^2}{m^2} + \frac{11}{9} \right) \frac{\bg{\sigma}}{2} + \frac{1}{4m^2} \frac{2}{3}  (\bg{\sigma} \times \ma{p}) \times \ma{p} \right\}u_h \ ,
\end{eqnarray}
where all IR divergences cancel. The {\it non-local} contributions are
\begin{eqnarray}
\label{2body1loopOrbital}
  \ma{f}(p,p)^{(1)}_{L_q, {\rm non-local}} & = & - \ma{f}(p,p)^{(1)}_{J_\gamma, {\rm non-local}} = \frac{\alpha_{\rm _{EM}}}{2 \pi} \left( -\frac{4}{3\epsilon_{\rm UV}} - \frac{4}{3}\ln\frac{\mu^2}{m^2} - \frac{17}{9} \right)u_h^\dag \ma{p}u_h \ .
\end{eqnarray}
As a cross check, we sum up the one-loop results
\begin{eqnarray}
\la \ma{S}_q + \ma{L}_q + \ma{J}_\gamma \ra^{(1)}_{\rm local} &=& 0\nonumber\\
\ma{f}(p,p)^{(1)}_{L_q, {\rm non-local}} + \ma{f}(p,p)^{(1)}_{J_\gamma, {\rm non-local}} &=& 0 \ .
\end{eqnarray}
The vanishing sum of the first-order contributions is expected since the total angular momentum is a conserved quantity. The algebra it respects dictates that the expectation value must be integer or half-integer and should not depend on expansion parameter $\alpha_{\rm _{EM}}$.


For the NRQED diagrams, we use dimensional regularization as well. Following the argument in Ref.~\cite{Manohar:1997qy}, in DR all loop diagrams in NRQED are zero due to scaleless integrals. The lack of scale is a natural consequence of the multi-pole expansion~\cite{Labelle:1996en}.  In the one-loop diagram containing ultrasoft photons in NRQED, the mass $m$ decoupled and the relevant scales are IR cutoff and UV cutoffs in the effective theory. However, in DR both divergences are regulated by the extra dimension $4-D$ and the only scale is $\mu$. Hence, all the NRQED loop integrals take the form
\begin{equation}
w_i^{(0)} \cdot \la \m{O}_{i,\rm eff}\ra^{(1)}(\mu) = \frac{\alpha_{\rm _{EM}}}{2\pi} A_{\rm eff}\left(\frac{1}{\epsilon_{\rm UV}}-\frac{1}{\epsilon_{\rm IR}}\right) \ .
\end{equation}
where $w_i^{(0)}$ is the zero-th order Wilson coefficient and $\la \m{O}_{i,\rm eff}\ra^{(1)}$ is the first-order matrix element of effective operators, which is scale-independent.

For a general matrix element in QED, we have
\begin{equation}
\label{IR_in_QED}
  \la\m{O}_{\rm QED} \ra^{(1)}= \frac{\alpha_{\rm _{EM}}}{2\pi}\left(\frac{A}{\epsilon_{\rm UV}} + \frac{B}{\epsilon_{\rm IR}} + (A+B)\ln\frac{\mu^2}{m^2} + C\right) \ .
\end{equation}
where UV divergence reflects the renormalization property of the operator in the full theory.
By imposing the matching condition,
\begin{eqnarray}
\label{IR_in_Eff}
  \la\m{O}_{\rm NRQED} \ra^{(1)}(\mu) & = & \la w_i\m{O}_{i,\rm eff} \ra^{(1)}(\mu) = w_i^{(0)} \cdot \la \m{O}_{i,\rm eff} \ra^{(1)}(\mu) + w_i^{(1)} \cdot \la \m{O}_{i,\rm eff} \ra^{(0)}(\mu) \nonumber\\
  & = & \frac{\alpha_{\rm _{EM}}}{2\pi} A_{\rm eff}\left(\frac{1}{\epsilon_{\rm UV}}-\frac{1}{\epsilon_{\rm IR}}\right) + w_i^{(1)}(\mu)\la \m{O}_{i,\rm eff} \ra^{(0)} .
\end{eqnarray}
Since the full and effective theories have to reproduce the same IR divergences, matching yields,
\begin{eqnarray}
&& A_{\rm eff} = -B, \, \, \, \, \,w_i^{(1)}(\mu) \la \m{O}_{i,\rm eff} \ra^{(0)} = \frac{\alpha_{\rm _{EM}}}{2\pi} (A +B)\ln\frac{\mu^2}{m^2} + C \ .
\end{eqnarray}
Threrefore, the Wilson coefficients can be obtained by subtracting all the $1/\e$ poles while keeping logarithms and finite terms in full theory calculation. Effectively this is equivalent to ``pulling up'' the IR divergences to the UV divergences in the Wilson coefficients.
Of course, this does not make sense since the Wilson coefficients should not depend on physics in the infrared.
All the IR sensitivities that appear in the full theory calculation should be reproduced by the matrix elements of the effective operators.
A serious problem resulting from the above procedure is the obscure physical meaning of the cutoff scale $\mu$ in the dimensional regularization.
It yields the wrong scaling behavior of the matrix elements $\frac{d \la \hat{\mathcal{O}}_J\ra}{d \ln\mu}$, where $\hat{\mathcal{O}}_J=\ma{S}_q$, $\ma{L}_q$ and $\ma{J}_\gamma$.
To alleviate this confusion in QED, one option is to regularize the IR divergence using a different regulator, e.g., the photon mass.
In the effective theory, the scaling behavior of the matrix elements also fails to satisfy the correct evolution equations
from the naive matching in dimensional regularization.
In Subsection III.F, we will render this problem by regularizing the UV divergence using a three-momenta cutoff $\Lambda$ in NRQED, which is distinguished from the QED scale $\mu$.
We will also use the photon mass to regularize the IR divergences.

At the moment, for the sake of simplicity, we use DR.

In NRQED, the contributions to the electron two-body matrix elements can be readily obtained from the operators in
Eqs.~(\ref{spin})-(\ref{gamma}). The {\it local} parts of matrix elements are
\begin{eqnarray}
  \la \ma{S}_q^{\rm eff} \ra_{\rm local} & = &  u_h^\dag\left\{ a_\sigma\frac{\bg{\sigma}}{2} + \frac{a_\pi}{4m^2} (\bg{\sigma}\times\ma{p}) \times\ma{p} \right\}u_h \ , \nonumber\\
  \la \ma{L}_q^{\rm eff} \ra_{\rm local} & = &  u_h^\dag\left\{ d_\sigma\frac{\bg{\sigma}}{2} + \frac{d_\pi}{4m^2} (\bg{\sigma}\times\ma{p}) \times\ma{p} \right\}u_h \ , \nonumber\\
  \la \ma{J}_\gamma^{\rm eff} \ra_{\rm local} & = &  u_h^\dag\left\{ f_\sigma\frac{\bg{\sigma}}{2} + \frac{f_\pi}{4m^2} (\bg{\sigma}\times\ma{p}) \times\ma{p} \right\}u_h  \ ,
\end{eqnarray}
and the {\it non-local} parts of matrix elements are
\begin{eqnarray}
  \ma{f}(p,p)_{L_q^{\rm eff}, {\rm non-local}} & = & d_R  u_h^\dag \ma{p} u_h \ , \nonumber\\
  \ma{f}(p,p)_{J_\gamma^{\rm eff}, {\rm non-local}} & = & f_R u_h^\dag  \ma{p} u_h \ .
\end{eqnarray}
Comparing them with Eqs.~(\ref{2bodytree}), (\ref{2body1loopSpin}) and (\ref{2body1loopOrbital}), we immediately arrive at some of the Wilson coefficients in Eq.~(\ref{spin})--(\ref{gamma}) in NRQED
\begin{eqnarray}
  a_\sigma&=&1+\frac{\alpha_{\rm _{EM}}}{2\pi},\;\;a_\pi=1+\frac{\alpha_{\rm _{EM}}}{2\pi} \ , \nonumber \\
  d_R &=& 1 + \frac{\alpha_{\rm _{EM}}}{2\pi}\left(-\frac{4}{3}\ln\frac{\mu^2}{m^2}-\frac{17}{9}\right),\;\; d_\sigma = \frac{\alpha_{\rm _{EM}}}{2\pi}\left(-\frac{4}{3}\ln\frac{\mu^2}{m^2}-\frac{20}{9}\right),\;\;d_\pi = -1 - \frac{5\alpha_{\rm _{EM}}}{6\pi} \ ,\nonumber \\
  f_R &=& \frac{\alpha_{\rm _{EM}}}{2\pi}\left(\frac{4}{3}\ln\frac{\mu^2}{m^2}+\frac{17}{9}\right),\;\; f_\sigma = \frac{\alpha_{\rm _{EM}}}{2\pi}\left(\frac{4}{3}\ln\frac{\mu^2}{m^2}+\frac{11}{9}\right),\;\;f_\pi = \frac{\alpha_{\rm _{EM}}}{3\pi} \ .
\end{eqnarray}
These results agree with Eq.~(7) of Ref.~\cite{Chen:2009rw}.
Again, this result assumes dimensional regularization for IR and UV divergences in the effective theory calculations.

\section{Three-Body Matching Through Electron Scattering in Background Field}

The two-body matching is insufficient to derive all the Wilson coefficients for the effective operators.
For example, in two-body calculations, these operators involving at least one photon field
give null result.
To obtain the complete expansion, we have to consider more complicated matrix elements.
In this section, we consider the processes with an electron interacting with an external electromagnetic field,
through the angular momentum operators. Together with the results from the previous section,
we will be able to determine the Wilson coefficients for all effective operators
bilinear in the electron fields.

\subsection{Tree Level}
We calculate the QED amplitude $\la e_{p'}|\m{O}_J|e_{p} \ma{A}_{q}\ra$ with an external vector potential $\ma{A}$, as well as $ \la e_{p'}|\m{O}_J|e_{p} A^0_{q}\ra$ with an external scalar potential $A^0$. The in- and out-electron states are always taken
to be on shell. We further assume electrons are non-relativistic and the virtual photons carry a small momentum $q$ compared
with the electron mass, i.e., $\ma{q}^2\ll m^2$.

\begin{figure}[hbs]
\begin{center}
\includegraphics[width=5cm]{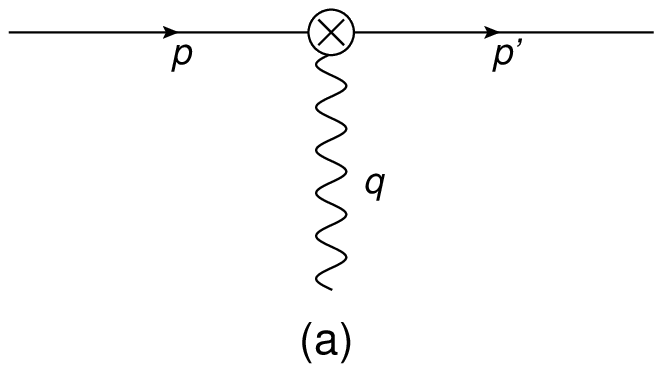}
\hspace{1cm}
\includegraphics[width=5cm]{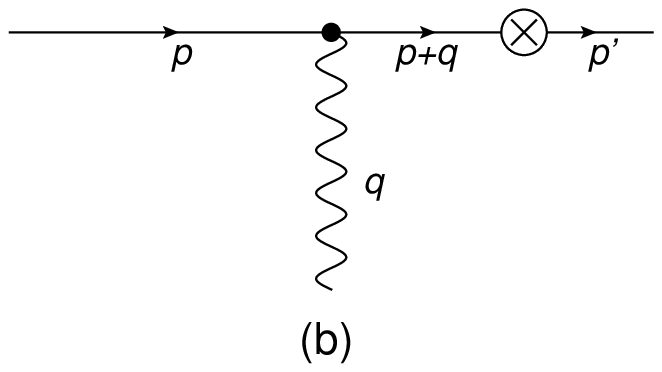}
\caption{QED tree diagrams for electron scattering in a photon background, coupled to angular momentum operators.}
\label{QED_Tree}
\end{center}
\end{figure}

In QED, the tree-level diagrams for the three-body matrix elements are shown in Fig.~\ref{QED_Tree}.
Fig.~\ref{QED_Tree} (a) is the direct contribution is from the $\ma{x}\times\ma{A}$ part in $\ma{L}_q$. Fig.~\ref{QED_Tree}
(b) is an indirect contribution.
We calculate the elastic scattering matrix element for on-shell electron $p^0=p'^0=\mathbf{p}^2+m^2$, $|\ma{p}| = |\ma{p'}|$.
The incident photon momentum is $q^\mu=(0,\ma{q})$, in which $\ma{q}=\ma{p}'-\ma{p}$.

Clearly the diagrams such as Fig.~\ref{QED_Tree}(b) will be
divergent if we take $p+q=p'$ since the intermediate state carries a
momentum on mass shell. This is a pole term. During the
matching, part of this pole term will be exactly reproduced by the non-relativistic electron
propagator in NRQED. There is also a finite contribution though,
due to the propagation of positron mode.
According to Eq.~(\ref{nr}), the virtual position propagation will be far off-shell and suppressed by $\m{O}(m^{-1})$.
Effectively, the propagator of Fig.~\ref{QED_Tree}(b) will shrink to a point and result in a contribution from
a local effective operators like those in Fig.~\ref{QED_Tree}(a).
The physics beyond the scale $m$ shall be be in the corresponding Wilson coefficients.

To regulate the above pole term, we introduce a small off-shellness to the intermediate electron
by assigning the photon momentum $q^\mu$ a time-like component, $q^\mu\rightarrow(q^0,\mathbf{q})$.
Meanwhile we rewrite the intermediate angular momentum operators as
$\hat{\m{O}}\rightarrow\hat{\m{O}}e^{iq^0 x^0}$ to maintain the four-momentum conservation.
Diagrammatically, we have an additional small momentum $(q^0,0)$ flowing in from the photon and out
from the angular momentum operators. Eventually, we will take $q^0\rightarrow0$.

For the matrix elements, we separate them into the $1/q^0$ pole term and the regular part in $q^0$, which represent the contribution from non-local electron propagation and the local interactions in NRQED, respectively.
In practice, it is enough to match only the local part of the matrix elements between QED and NRQED.
Matching the terms carrying $1/q^0$ poles can be used as a consistency check of our results.

In the presence of $q^0$, the denominator of the intermediate electron propagator in Fig.~\ref{QED_Tree}(b) can be Taylor expanded as~\cite{Luke:1999kz}
\begin{eqnarray}
\frac{1}{(p+q)^2-m^2} & = & \frac{1}{2m q^0} - \frac{1}{4m^2} + \frac{q^0}{8m^3} - \frac{\ma{p}^2}{4m^3q^0} + \cdots \ ,
\end{eqnarray}
where we have used the on-shell kinematics for the electron $p^2=p'^2=m^2$, $\mathbf{p}' = \mathbf{p} + \mathbf{q}$ and $p'^0=p^0\simeq m + \ma{p}^2/2m$
and make subsequent non-relativistic expansion.
Later on we shall make the same expansion in powers of $q^0$ in the next-to-leading-order calculations as well.

For non-local operators, we choose $\mathbf{x}\rightarrow i\bg{\nabla}_q$,
i.e. the derivative with respect to the external momentum of the photon. This has several advantages.
First, the result will be symmetric with respect to the incoming and outgoing electron momenta.
Second, we need not consider the derivative on the photon polarization which does not depend on $\ma{q}$.
This greatly simplifies the calculation.
Finally, the three-momentum conservation relation $\mathbf{p'}=\mathbf{p}+\mathbf{q}$ should be understood
after the replacement $\mathbf{x}\rightarrow i\bg{\nabla}_q$ has been made.

The tree-level non-pole contributions from Fig.~\ref{QED_Tree} in QED are
\begin{eqnarray}
\label{QED_Tree_Amplitude}
&&\la \mathbf{S}_q \ra_{\rm local}^{(0)} = \frac{e}{4m^2}u_h^\dag \left[i\ma{q}\times\ma{A} + (\mathbf{P}\cdot\mathbf{A})\bg{\sigma} - (\mathbf{P}\cdot\bg{\sigma})\mathbf{A}\right]u_h \ ,\nonumber\\
&&\la \mathbf{L}_q \ra_{\rm local}^{(0)} = \frac{e}{4m^2}u_h^\dag \left[-i\ma{q}\times\ma{A} - (\mathbf{P}\cdot\mathbf{A})\bg{\sigma} + (\mathbf{P}\cdot\bg{\sigma})\mathbf{A}\right]u_h \ ,\nonumber\\
&&\ma{f}(p',p,p'-p)^{(0)}_{L_q, {\rm non-local}} = u_h^\dag \left[1 - \frac{|\mathbf{q}|^2}{8m^2} + \frac{i}{8m^2} (\mathbf{q}\times\mathbf{P})\cdot\bg{\sigma}\right](-e\mathbf{A})u_h \ ,
\end{eqnarray}
in which $\ma{P} \equiv \ma{p}'+\ma{p}$ and $\ma{q} \equiv \ma{p}' - \ma{p}$. Terms of order $\m{O}(m^{-3})$ have been suppressed.

\begin{figure}[!htb]
\begin{center}
\includegraphics[width=4cm]{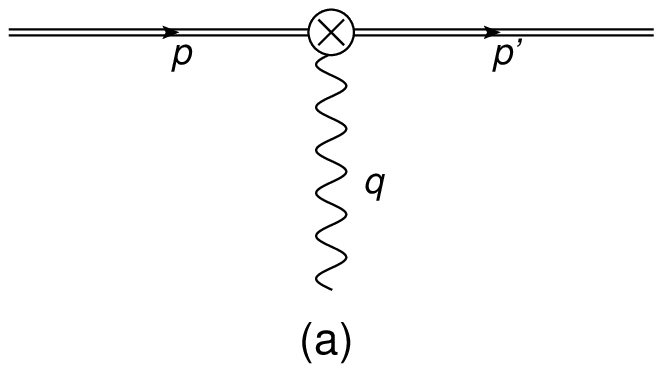}
\hspace{.5cm}
\includegraphics[width=4cm]{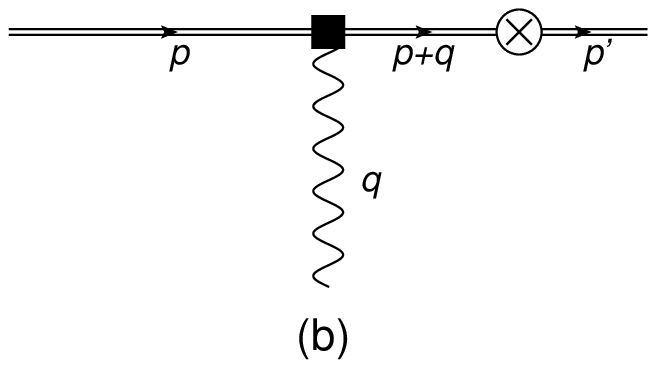}
\hspace{.5cm}
\includegraphics[width=4cm]{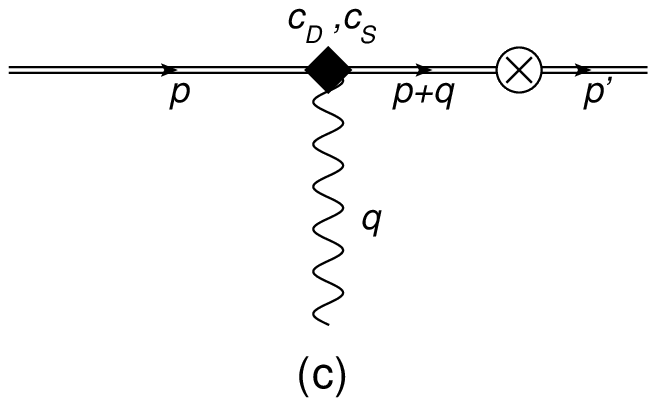}
\caption{corresponding NRQED tree diagrams as in Fig. 2}
\label{NR_Tree}
\end{center}
\end{figure}

Next, we consider the tree diagrams in NRQED as shown in Fig.~\ref{NR_Tree}. The electron propagator in Fig.~\ref{NR_Tree}(b),(c)
\begin{equation}
  \frac{1}{E_p - \frac{(\ma{p}+\ma{q})^2}{2m} + i\epsilon}\to\frac{1}{E_{p'} - \frac{\ma{p}'^2}{2m}+i\epsilon} \ ,
\end{equation}
is also divergent for on-shell electron external states. By introducing a small non-zero $q^0$, 
the fermion propagator will be regularized to $\frac{i}{q^0+i\epsilon}$. 
Since there is no angular momentum effective operator containing time derivatives,
the $1/q^0$ pole can only be canceled by Dirac and spin-orbit terms in the NRQED Lagrangian 
involving the electric field strength $\ma{E}$\ (since $\ma{E}^i = -F^{0i} \sim i(q^0\ma{A}^i-q^iA^0)$ contains positive powers of $q^0$), 
resulting in local matrix elements.

As a concrete example, we calculate the {\it local} part of the matrix element $\la e|\int d^3\ma{x}\psi^\dag\ma{x}\times\bg{\pi}\psi|e\ma{A}\ra^{\rm eff,Tree}_{\rm local}$ as shown in Fig.~\ref{NR_Tree},
\begin{eqnarray}\label{eg}
  \la e|\ma{x}\times\bg{\pi}|e\ma{A}\ra^{(0)}_{\rm local} & = & i\bg{\nabla}_q\times u_h^\dag \left\{(-e\ma{A}) + \left(\frac{i\ma{p}}{q^0+i\epsilon} + \frac{i\ma{p}'}{- q^0 +i\epsilon}\right) \cdot \frac{ie c_D}{8m^2}\ma{q}\cdot(\ma{q}A^0 - q^0\ma{A}) \right.\nonumber\\
  &+& \frac{ec_S}{8m^2}q^0\left(\left[(2\ma{p}'-\ma{q}) \cdot \ma{A} \times \bg{\sigma}\right]\frac{i\ma{p}} {q^0+i\epsilon} + \left[(2\ma{p} + \ma{q})\cdot \ma{A} \times \bg{\sigma}\right] \frac{i\ma{p}'}{-q^0 +i\epsilon} \right) \nonumber\\
  &+&\left. \frac{ec_S}{8m^2}A^0\ma{\sigma}\cdot(\ma{P}\times\ma{q})\cdot \left(\frac{i\ma{p}}{q^0+i\epsilon} + \frac{i\ma{p}'}{- q^0 +i\epsilon}\right) \right\}u_h \nonumber\\
  & \sim & u_h^\dag\left[-\frac{\ma{p'}-\ma{p}}{q^0} \times \frac{ie c_D}{8m^2} q^0 \ma{A} + \frac{\ma{p} + \ma{p}'}{q^0}\times \frac{e c_S}{8m^2}q^0(\ma{A}\times\bg{\sigma}) \right]u_h  \nonumber\\
  & = & u_h^\dag\left[-\frac{iec_D}{8m^2} \ma{q}\times\ma{A} - \frac{ec_S}{8m^2}\ma{P} \times (\bg{\sigma}\times \ma{A})\right]u_h  \ .
\end{eqnarray}
In the above calculation, we have included both cases that the NRQED interaction happens before and after
the angular momentum operator, along the momentum flow.
As promised above, we only need to keep the non-pole terms [regular as $q^0\to 0$] and suppress all the  $1/q^0$ poles.

Following the same procedure, we have calculate non-pole contributions from the local and non-local
matrix elements for all the effective operators in Eqs.~(\ref{spin})-(\ref{gamma}). The results are listed in Table I.

\begin{table}[!htb]
\begin{center}
\label{NRQED_BiQuark}
\begin{tabular}
{|c|c|c|}\hline
 {\rm operator} & {\it local} & {\it non-local} \\
 \hline\hline
 $ \psi^\dag (\bg{\sigma}/2) \psi$ & $-\frac{e}{8m^2}c_S u_h^\dag (\ma{P}\times\ma{A})\times\bg{\sigma} u_h$ & -\\
 \hline
 $ \psi^\dag \ma{x}\times\bg{\pi}\psi $ & $-\frac{e}{8m^2}u_h^\dag\left[c_D  (i\ma{q}\times\ma{A}) \right.$ &  $\frac{e}{8m^2} u_h^\dag\left[c_D (\ma{q}\cdot\ma{A})\ma{q}  \right.$\\
  & $\left. + c_S \ma{P}\times(\bg{\sigma}\times \ma{A})\right]u_h $& $\left. + ic_S\bg{\sigma}\cdot(\ma{P}\times\ma{A}) \ma{q}\right]u_h$\\
 \hline
 $ \psi^\dag\left[{\mathbf x} \times (\bg{\sigma} \times
\mathbf{E})\right]\psi $ & $2u_h^\dag(A^0\bg{\sigma})u_h $ & $-i u_h^\dag (A^0\bg{\sigma}\times\ma{q}) u_h$\\
 \hline
 $\psi^\dag \left[(\bg{\sigma}\times\bg{\pi})
\times \bg{\pi}\right. $ & $e u_h^\dag \left[ \ma{P}\times(\bg{\sigma}\times\ma{A})\right.$ & - \\
  $\left.-\bg{\pi} \times (\bg{\sigma}\times\bg{\pi})\right]\psi$ & $\left.+ \ma{A}\times(\bg{\sigma}\times\ma{P})\right] u_h$ & \\
\hline
 $\psi^\dag \ma{B}\psi$ & $iu_h^\dag \ma{q}\times\ma{A} u_h$ & -\\
\hline
 $ {\mathbf x}\times \left[\mathbf B \times \bg{\nabla}(\psi^\dag\psi)\right] $ & $i u_h^\dag \ma{q}\times\ma{A} u_h$ & $u_h^\dag \left[\ma{q}^2\ma{A} - (\ma{q}\cdot\ma{A})\ma{q}\right]u_h$\\
 \hline
 $ {\mathbf x}\times \psi^\dag \left[\mathbf{B} \times(\bg{\sigma} \times \overleftrightarrow{\bg{\pi}}) \right]\psi $ & $u_h^\dag \ma{A}\times(\bg{\sigma}\times\ma{P})u_h$ & $-i u_h^\dag\left[ \bg{\sigma}\cdot(\ma{q}\times\ma{P})\ma{A} \right.$ \\
  & & $\left. + \bg{\sigma}\cdot(\ma{P}\times\ma{A})\ma{q}\right]u_h$\\
 \hline
 ${\mathbf x}\times \psi^\dag \left[\bg{\sigma}\times (\mathbf{B} \times
\overleftrightarrow{\bg{\pi}}) \right]\psi$ & $u_h^\dag \left[(\bf{\sigma}\cdot\ma{P})\ma{A} + (\ma{A}\cdot\ma{P})\bg{\sigma}\right]u_h$ & $i u_h^\dag\left[(\ma{q}\cdot\ma{P})\bg{\sigma}\times\ma{A}\right.$\\
 & & $\left. - (\ma{A}\cdot\ma{P})\bg{\sigma}\times\ma{q}\right]u_h$\\
\hline
\end{tabular}
\caption{{\it Local} and {\it non-local} contributions to $\la e|\m{O}^{\rm eff}|e\mathbf{B}\ra_{\rm Tree}$, only local terms are shown.}
\end{center}
\end{table}

From Eqs.~(\ref{spin}), (\ref{orbit}) and (\ref{QED_Tree_Amplitude}), we get the leading order Wilson coefficients up to $\mathcal{O}(1)$
\begin{eqnarray}
&  a_\sigma^{(0)} = a_\pi^{(0)} = a_B^{(0)} = 1\ ,&\nonumber\\
&  d_\sigma^{(0)} = -d_\pi^{(0)} = -d_B^{(0)} = d_D^{(0)} = d_S^{(0)} = 1& \ .
\end{eqnarray}
They agree with the non-relativistic reductions through the FW transformation in Eq.~(\ref{FW}).
The other Wilson coefficients in $\ma{S}_q^{\rm eff}$ and $\ma{L}_q^{\rm eff}$ are of order $\m{O}(\alpha_{\rm _{EM}})$ or higher.
They will show up in one-loop matching as described in the next subsection.

\subsection{One-Loop Matching}

\begin{figure}[!htb]
\begin{center}
\includegraphics[width=3cm]{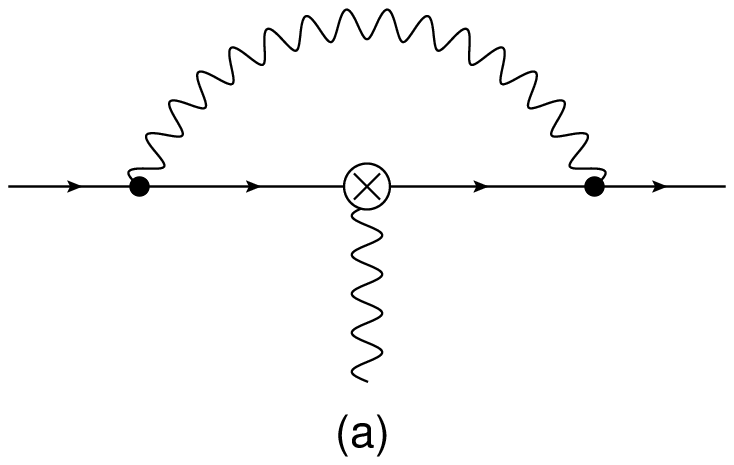}
\hspace{2mm}
\includegraphics[width=3cm]{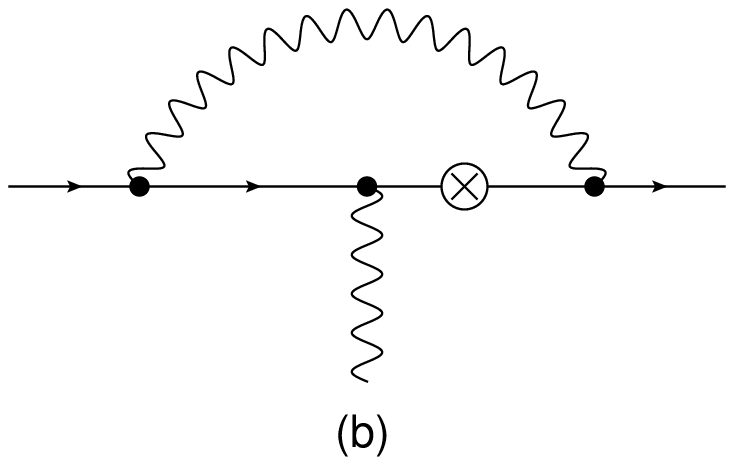}
\hspace{2mm}
\includegraphics[width=3cm]{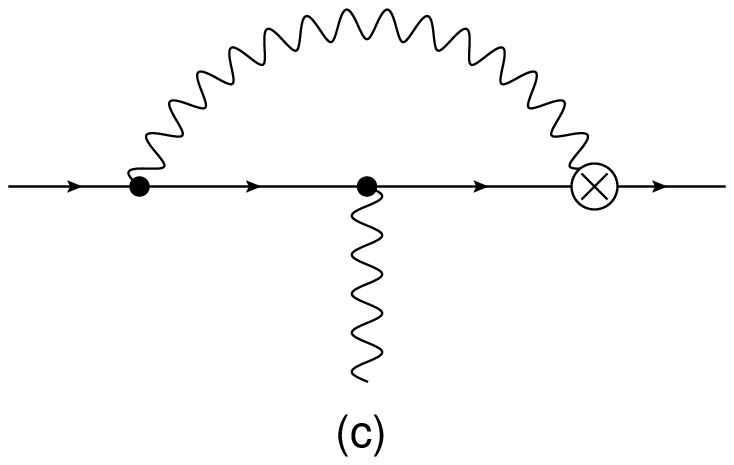}
\hspace{2mm}
\includegraphics[width=3cm]{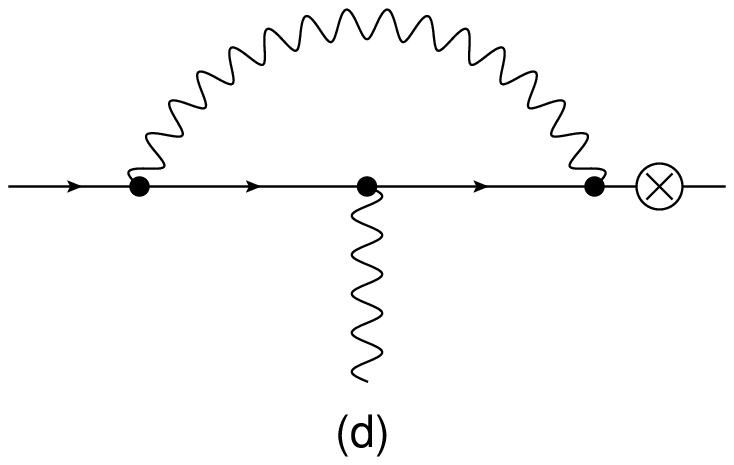}\\
\vspace{2mm}
\includegraphics[width=3cm]{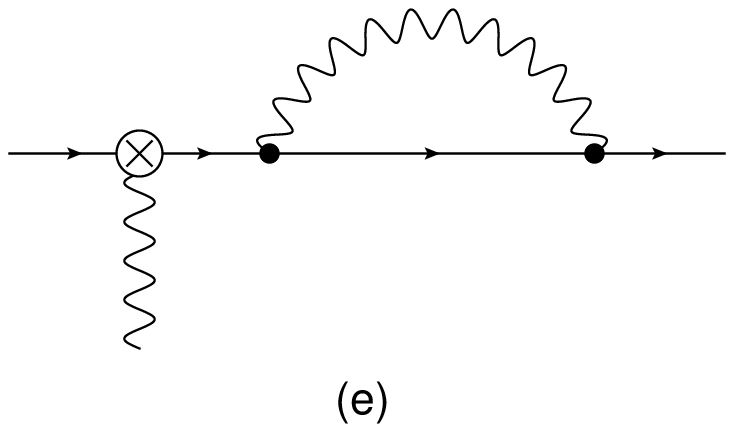}
\hspace{2mm}
\includegraphics[width=3cm]{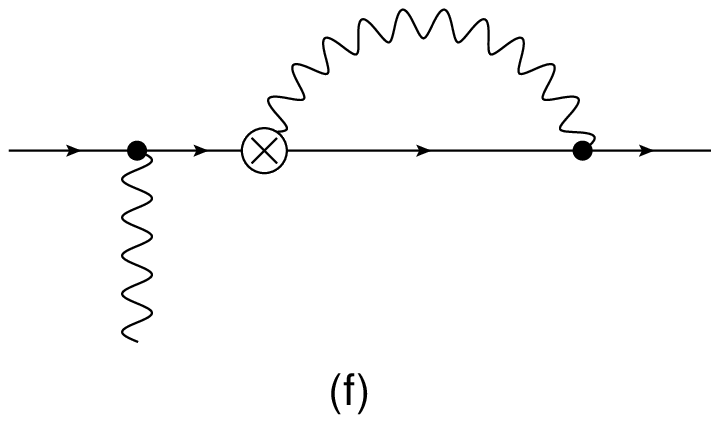}
\hspace{2mm}
\includegraphics[width=3cm]{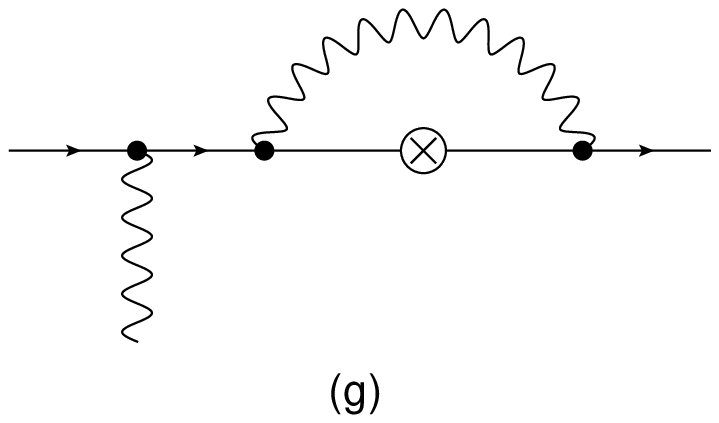}
\hspace{2mm}
\includegraphics[width=3cm]{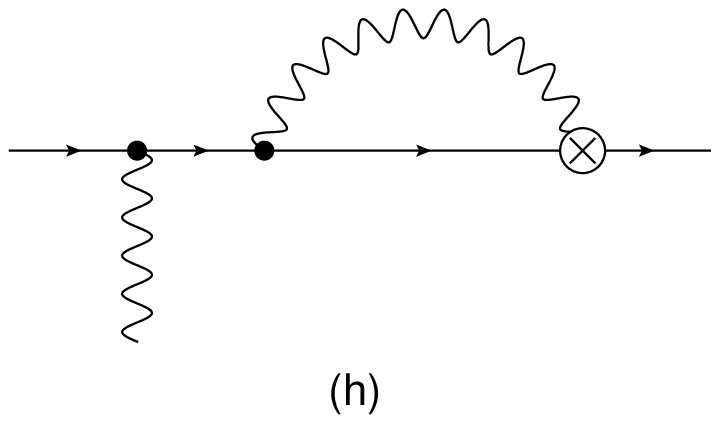}\\
\vspace{2mm}
\includegraphics[width=3cm]{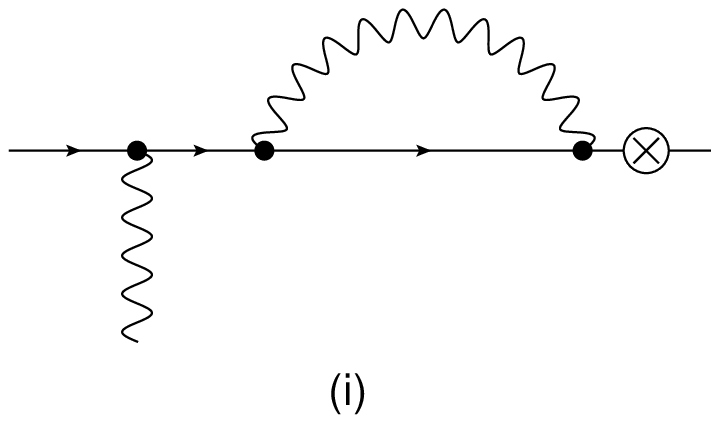}
\hspace{2mm}
\includegraphics[width=3cm]{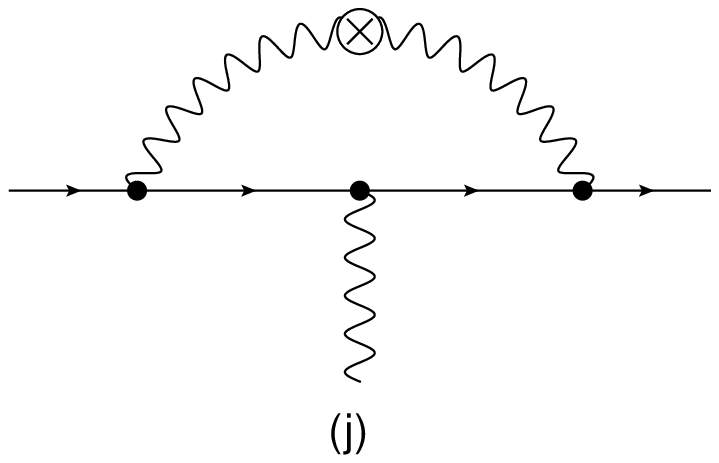}
\hspace{2mm}
\includegraphics[width=3cm]{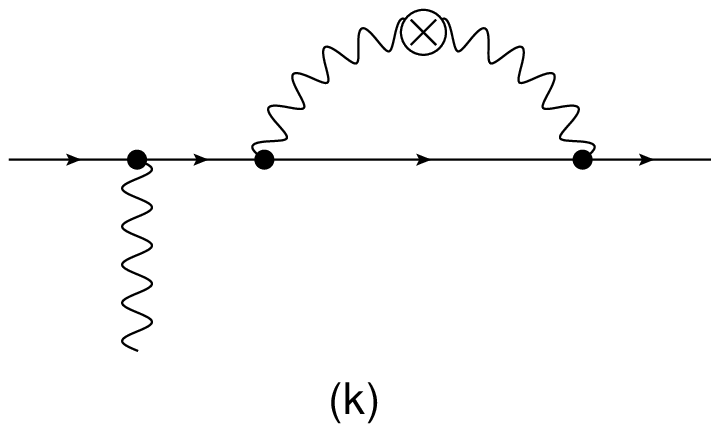}
\caption{One-loop contributions to $\la e|\mathbf{J}|e\gamma\ra$ in QED.  Here $\otimes$ can be either $\ma{L}_q$,$\ma{S}_q$, or $\ma{J}_\gamma$: (a)-(i) for $\ma{L}_q$; (b)(d)(g)(i) for $\ma{S}_q$; (j)(k) for $\ma{J}_\gamma.$  Wave function renormalization diagrams, mass counter terms and the mirror diagrams are not shown explicitly. }
\label{QED_OneLoop}
\end{center}
\end{figure}

In this subsection, we calculate the one-loop radiative corrections to angular momentum operators in QED and NRQED.
To find the complete set of Feynman diagrams, we first draw the one-loop QED diagrams and insert angular momentum
operators in all possible ways. The corresponding Feynman diagrams in QED are shown in Fig.~\ref{QED_OneLoop}. Here
we have no diagram with $\ma{J}_\gamma$ on the external photon line because this yields only pole contribution.
The photon pole contribution is related to matching in the photon state, which will be done
in the next subsection.

The loop diagrams are considerably more complicated than the
tree-level case. This is because all loops can be Taylor expanded in
$q^0$ and contribute to the local matrix elements in a similar
manner as Eq.~(\ref{eg}). Again only the regular part of matrix
elements in the limit $q^0\rightarrow 0$ is taken into account.

After lengthy calculations we find, for the vector external potential $\ma{A}$ case, the {\it local} part of matrix element reads
\begin{eqnarray}
\label{S_spin}
\la \mathbf{S}_q \ra^{(1)}_{\rm local} & = & \frac{\alpha_{\rm _{EM}} }{2\pi} \frac{e}{4m^2} u_h^\dag\left[7i\ma{q}\times\ma{A} + (\mathbf{P}\cdot\mathbf{A})\bg{\sigma} + (\bg{\sigma}\cdot\mathbf{A})\mathbf{P} - 2(\mathbf{P}\cdot\bg{\sigma})\mathbf{A}\right]u_h \ ,
\end{eqnarray}
\begin{eqnarray}
\label{L_spin}
\la \mathbf{L}_q \ra_{\rm local}^{(1)} & = & \frac{\alpha_{\rm _{EM}}}{2\pi}\frac{e}{4m^2}u_h^\dag\left[ \left(\frac{8}{\e_{\rm IR}} + 8 \ln\frac{\mu^2}{m^2} + \frac{50}{3}\right)i\ma{q}\times\ma{A} + \frac{2}{3}(\mathbf{P}\cdot\mathbf{A})\bg{\sigma} - 2(\bg{\sigma}\cdot\mathbf{A})\mathbf{P} + \frac{8}{3}(\mathbf{P}\cdot\bg{\sigma})\mathbf{A}\right]u_h \ , \nonumber \\
\end{eqnarray}
\begin{eqnarray}
\label{J_spin}
\la \mathbf{J}_\gamma \ra_{\rm local}^{(1)} & = & \frac{\alpha_{\rm _{EM}}}{2\pi}\frac{e}{4m^2}u_h^\dag\left[ \left(-\frac{8}{\e_{\rm IR}} - 8 \ln\frac{\mu^2}{m^2} - \frac{71}{3}\right)i\ma{q}\times\ma{A} - \frac{5}{3}(\mathbf{P}\cdot\mathbf{A})\bg{\sigma} + (\bg{\sigma}\cdot\mathbf{A})\mathbf{P} - \frac{2}{3}(\mathbf{P}\cdot\bg{\sigma})\mathbf{A}\right]u_h \ , \nonumber \\
\end{eqnarray}
and the {\it non-local} part
\begin{eqnarray}
\label{L_orbital}
\ma{f}(p',p,p'-p)^{(1)}_{L_q, {\rm non-local}} & = & \frac{\alpha_{\rm _{EM}}}{2\pi}\frac{e}{4m^2}u_h^\dag\left[ \left(\frac{4}{3\e_{\rm UV}} + \frac{4}{3}\ln\frac{\mu^2}{m^2} + \frac{17}{9}\right)(4m^2)\mathbf{A} - \frac{2}{3}i (\mathbf{P}\cdot\mathbf{A}) (\bg{\sigma} \times \mathbf{q})\right.\nonumber\\
 & &  - \frac{5}{3}i [\bg{\sigma}\cdot(\mathbf{P}\times\mathbf{A})] \mathbf{q} + \left(\frac{2}{3\e_{\rm UV}} + \frac{2}{3}\ln\frac{\mu^2}{m^2} - \frac{31}{18}\right)i [\bg{\sigma}\cdot(\mathbf{q}\times\mathbf{P})] \mathbf{A} \nonumber\\
 & & \left.  - \frac{7}{9}(\mathbf{q}\cdot\mathbf{A})\mathbf{q} + \left( -\frac{2}{3\e_{\rm_{UV}}} - \frac{4}{3\e_{\rm_{IR}}} -2 \ln\frac{\mu^2}{m^2} - \frac{1}{6} \right)|\mathbf{q}|^2\mathbf{A} \right]u_h \ ,
\end{eqnarray}
\begin{eqnarray}
\label{g_orbital}
\ma{f}(p',p,p'-p)^{(1)}_{J_\gamma, {\rm non-local}} & = & \frac{\alpha_{\rm _{EM}}}{2\pi}\frac{e}{4m^2} u_h^\dag\left[ \left(-\frac{4}{3\e_{\rm UV}}-\frac{4}{3}\ln\frac{\mu^2}{m^2} - \frac{17}{9}\right)(4m^2)\mathbf{A} + \frac{2}{3}i (\mathbf{P}\cdot\mathbf{A}) (\bg{\sigma} \times \mathbf{q}) \right.\nonumber\\
 & & + \frac{5}{3}i [\bg{\sigma}\cdot(\mathbf{P}\times\mathbf{A})] \mathbf{q} + \left(-\frac{2}{3\e_{\rm UV}} - \frac{2}{3}\ln\frac{\mu^2}{m^2} + \frac{13}{18}\right)i [\bg{\sigma}\cdot(\mathbf{q}\times\mathbf{P})] \mathbf{A} \nonumber\\
 & & \left. + \frac{7}{9}(\mathbf{q}\cdot\mathbf{A})\mathbf{q}  + \left( \frac{2}{3\e_{\rm UV}} + \frac{2}{3}\ln\frac{\mu^2}{m^2} + \frac{1}{6} \right)|\mathbf{q}|^2\mathbf{A} \right]u_h  \ .
\end{eqnarray}

For the scalar potential $A^0$ case, the local part of matrix elements are
\begin{eqnarray}
  \la \mathbf{S}_q \ra_{\rm local}^{(1)} & = & \m{O}(m^{-3}) \ ,\nonumber\\
  \la \mathbf{L}_q \ra_{\rm local}^{(1)} & = & \frac{\alpha_{\rm _{EM}}}{2\pi}\frac{e}{4m^2} u_h^\dag\left[\frac{4m}{3}A^0 \bg{\sigma}\right] u_h + \m{O}(m^{-3}) \ ,\nonumber\\
  \la \mathbf{J}_\gamma \ra_{\rm local}^{(1)} & = & \frac{\alpha_{\rm _{EM}}}{2\pi}\frac{e}{4m^2}u_h^\dag\left[ -\frac{4m}{3}A^0\bg{\sigma} \right]u_h + \m{O}(m^{-3}) \ ,
\end{eqnarray}
and the {\it non-local} part
\begin{eqnarray}
  \ma{f}(p',p,p'-p)^{(1)}_{L_q, {\rm non-local}} & = & \frac{\alpha_{\rm _{EM}}}{2\pi}\frac{e}{4m^2} u_h^\dag\left[ -\frac{2m}{3}A^0i(\bg{\sigma}\times\mathbf{q}) \right] u_h \ ,\nonumber\\
  \ma{f}(p',p,p'-p)^{(1)}_{J_\gamma, {\rm non-local}} & = & \frac{\alpha_{\rm _{EM}}}{2\pi}\frac{e}{4m^2} u_h^\dag\left[ \frac{2m}{3}A^0i(\bg{\sigma}\times\mathbf{q}) \right] u_h \ .
\end{eqnarray}

Again we sum up all {\it local} and {\it non-local} matrix elements of the angular momentum and get
\begin{eqnarray}
&&  \la e| \ma{S}_q + \ma{L}_q + \ma{J}_\gamma| e \ma{A}\ra_{\rm local}^{(1)} = \la e| \ma{S}_q + \ma{L}_q + \ma{J}_\gamma| e A^0\ra_{\rm local}^{(1)} = 0,\nonumber\\
&&  \la e| \ma{S}_q + \ma{L}_q + \ma{J}_\gamma| e \ma{A}\ra_{\rm non-local}^{(1)} = \frac{\alpha_{\rm _{EM}}}{2\pi}\frac{e}{4m^2} \mathbf{x} \times u_h^\dag\left[\rule{0mm}{5.5mm}-2i [\bg{\sigma}\cdot(\mathbf{q}\times\mathbf{P})] \mathbf{A} -\frac{4}{3} \left(\frac{1}{\e_{\rm IR}} +\ln\frac{\mu^2}{m^2}\right)|\ma{q}|^2\ma{A}  \right]u_h \ ,\nonumber\\
&&  \la e| \ma{S}_q + \ma{L}_q + \ma{J}_\gamma| e A^0\ra_{\rm non-local}^{(1)} = 0 \ .
\end{eqnarray}
The reason for the null results is again due to the conservation of the total angular momentum.
The non-zero first-order result for $\la e| \ma{J}_{\rm NRQED}| e \ma{A}\ra_{\rm orbital}^{(1)}$ is due to the non-vanishing one-loop contributions of the electromagnetic form factors $F_1(q^2)$ and $F_2(q^2)$. Once we include the diagrams with the operator $J_\gamma$ placed on the external photon line, the total orbital contribution to the matrix element $\la e| \ma{J}_\gamma| e \ma{A}\ra_{\rm orbital}^{(1)}$ would also yield zero.

For the matrix elements in NRQED, we only need to calculate the
tree-level diagrams for the effective operators in dimensional
regularization. By applying the matching conditions according to
Eq.~(\ref{MatchingCondition}) together with the results in Table~I,
all the Wilson coefficients for effective operators bilinear in
quark fields can be determined in the $\overline{\rm MS}$ scheme,
\begin{eqnarray}
\label{WilsonBilinear}
&&a_\sigma = 1 + \frac{\alpha_{\rm _{EM}}}{2\pi},\;\; a_\pi = 1 + \frac{\alpha_{\rm _{EM}}}{2\pi}, \;\; a_B = 1 + \frac{7\alpha_{\rm _{EM}}}{2\pi},\;\; \ ,\nonumber\\
&&d_R = 1 + \frac{\alpha_{\rm _{EM}}}{2\pi} \left(-\frac{4}{3}\ln\frac{\mu^2}{m^2} - \frac{17}{9}\right),\;\; d_\sigma = \frac{\alpha_{\rm _{EM}}}{2\pi}\left(-\frac{4}{3}\ln\frac{\mu^2}{m^2} - \frac{20}{9}\right),\;\; d_\pi = -1 - \frac{5\alpha_{\rm _{EM}}}{6\pi},\nonumber\\
&&d_D = 1 + \frac{\alpha_{\rm _{EM}}}{2\pi}\left(-4\ln\frac{\mu^2}{m^2} - \frac{1}{3}\right),\;\; d_S = 1 + \frac{\alpha_{\rm _{EM}}}{2\pi}\left(-\frac{4}{3}\ln\frac{\mu^2}{m^2} + \frac{31}{9}\right)\ ,\nonumber\\
&&d'_S = \frac{2\alpha_{\rm _{EM}}}{3\pi}, \;\; d_E = \frac{\alpha_{\rm _{EM}}}{3\pi}, \;\;
d_B = -1 + \frac{\alpha_{\rm _{EM}}}{2\pi} \left(8\ln\frac{\mu^2}{m^2} + \frac{143}{9}\right) \ ,\nonumber\\
&&f_R = \frac{\alpha_{\rm _{EM}}}{2\pi} \left(\frac{4}{3}\ln\frac{\mu^2}{m^2} + \frac{17}{9}\right),\;\; f_\sigma = \frac{\alpha_{\rm _{EM}}}{2\pi} \left(\frac{4}{3}\ln\frac{\mu^2}{m^2} + \frac{11}{9}\right),\;\;f_\pi = \frac{\alpha_{\rm _{EM}}}{3\pi},\nonumber\\
&&f_D = \frac{\alpha_{\rm _{EM}}}{2\pi}\left(\frac{4}{3}\ln\frac{\mu^2}{m^2} + \frac{1}{3}\right),\;\; f_S = - \frac{\alpha_{\rm _{EM}}}{2\pi}\left(-\frac{4}{3}\ln\frac{\mu^2}{m^2} + \frac{13}{9}\right),\nonumber\\
&&f'_S = -\frac{2\alpha_{\rm _{EM}}}{3\pi},\;\; f_E = -\frac{\alpha_{\rm _{EM}}}{3\pi}, \;\;
f_B = \frac{\alpha_{\rm _{EM}}}{2\pi}\left(-8\ln\frac{\mu^2}{m^2}-\frac{206}{9}\right)\ .
\end{eqnarray}
As it can be easily verified, the coefficients satisfy the sum rules in Eq. (20).

These results reproduce Eq.~(8) of Ref.~\cite{Chen:2009rw}, except for coefficients $d_D$, $d_B$ and $f_B$. We will properly separate the UV and IR divergences in these three coefficients when discussing the renormalization group evolution in Sec.\,VI.

\section{Photon Sector}

In this section, we calculate the forward matrix elements $\la\gamma|\hat{\m{O}}|\gamma\ra$ up to the
order $\m{O}(\alpha_{\rm _{EM}})$ with off-shell photons carrying a momentum $q$.
The tree-level matching is straightforward and we have
\begin{equation}
  f_\gamma^{(0)} = 1,\;\; d_\gamma^{(0)} = 0.
\end{equation}

\begin{figure}[hb]
\begin{center}
\includegraphics[width=3cm]{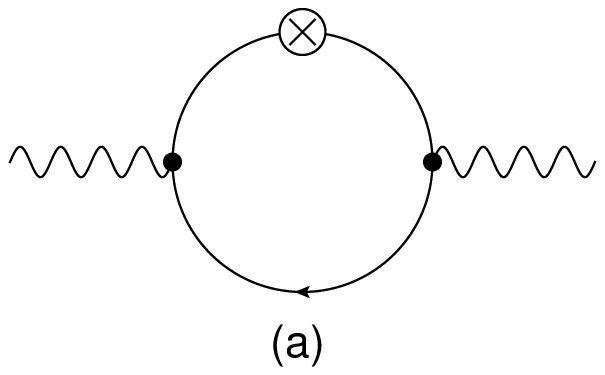}
\hspace{.5cm}
\includegraphics[width=3cm]{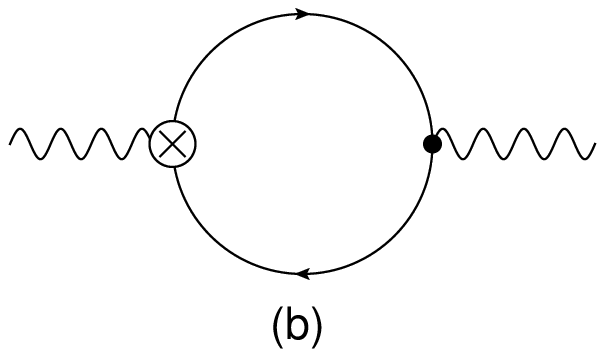}
\hspace{.5cm}
\includegraphics[width=3cm]{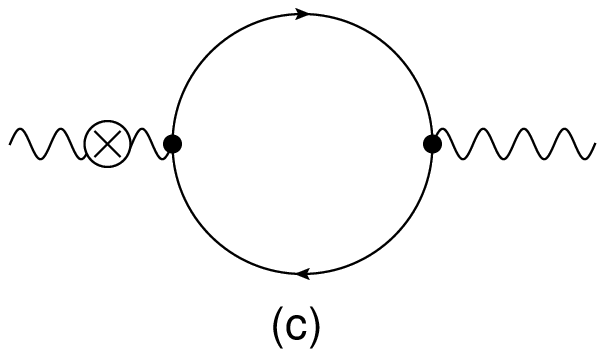}
\caption{One-loop contributions to $\la \gamma|\mathbf{J}|\gamma\ra$ in QED. Mirror diagrams are not shown. (a) for $\hat{S}_q$; (a)(b) for $\hat{L}_q$; (c) for $\hat{J}_\gamma$.}
\label{QED_photon}
\end{center}
\end{figure}

The relevant QED Feynman diagrams at one-loop are shown in Fig.~\ref{QED_photon}.
While the operators $\ma{S}_q$ and $\ma{L}_q$ can be inserted to the fermion loop of the vacuum polarization diagram as in Fig.~\ref{QED_photon}(a) and (b), the one-loop contribution from $\ma{J}_\gamma$ only arise in Fig.~\ref{QED_photon}(c). We need no new effective operator to match the matrix element of $\mathbf{J}_\gamma$, simply because the vacuum polarization is already reflected by the $d_1$- and $d_2$-terms in the effective Lagrangian. Thus
\begin{equation}
f_\gamma^{(1)} = 0 \ ,
\end{equation}
and there are no further corrections to $\ma{J}_\gamma$ at $\m{O}(\alpha_{\rm _{EM}}^0 m^{-2})$ order as already indicated in Eq.~(\ref{gamma}).

The calculation of the one-loop matrix elements for $\mathbf{S}_q$ and $\mathbf{L}_q$ in QED yields
\begin{eqnarray}
  \la\gamma |\mathbf{S}_q| \gamma\ra_{\rm local}^{(1)} & = & \frac{\alpha_{\rm _{EM}}}{2\pi}\frac{i}{m^2} \frac{2}{3} \left[ (\mathbf{A}\times\mathbf{A}^*)q^0q^2 - (\mathbf{A}\times\mathbf{q}) q^2A^{*0} -  (\mathbf{q}\times\mathbf{A}^*)q^2A^0\right] \ ,\\
  \la\gamma |\mathbf{L}_q| \gamma\ra_{\rm local}^{(1)} & = & \frac{\alpha_{\rm _{EM}}}{2\pi}\left(\frac{\mu^2}{m^2}\right)^\epsilon \frac{2i}{3\epsilon_{\rm UV}}\left[(\mathbf{A} \times \mathbf{A}^*)q^0 + (\mathbf{A}\times\mathbf{q})A^{*0} + 2 (\mathbf{q}\times\mathbf{A}^*)A^0 \right] \nonumber\\
  &+& \frac{\alpha_{\rm _{EM}}}{2\pi}\frac{i}{m^2}\frac{1}{15}\left[(\mathbf{A}\times\mathbf{A}^*)q^0q^2 + 3(\mathbf{A}\times\mathbf{q}) q^2A^{*0} + 13(\mathbf{q}\times\mathbf{A}^*)q^2A^0 \right.\nonumber\\
  &+& \left. 4(\mathbf{A}\times\mathbf{q})(q\cdot A^*)q^0 +4(\mathbf{q}\times\mathbf{A}^*)(q\cdot A)q^0 \right] \ , \\
  \ma{f}(q,q)^{(1)}_{L_q, {\rm non-local}} & = & \frac{\alpha_{\rm _{EM}}}{2\pi}\left(\frac{\mu^2}{m^2}\right)^\epsilon \frac{2}{3\epsilon_{\rm UV}} \left[-2(A \cdot A^*)q^0\mathbf{q} + (q\cdot A)q^0 \mathbf{A}^* + (q\cdot A^*)q^0 \mathbf{A} \right] \nonumber\\
  &+& \frac{\alpha_{\rm _{EM}}}{2\pi}\frac{1}{m^2} \frac{2}{15} \left[-4(A\cdot A^*)q^0q^2 \mathbf{q} + (q\cdot A)q^0q^2\mathbf{A}^* + (q\cdot A^*)q^0q^2\mathbf{A}\right. \nonumber\\
  &+& (q\cdot A)A^{*0}q^ 2\mathbf{q} + (q\cdot A^*)A^0q^2\mathbf{q} - (q^2)^2(A^0\mathbf{A}^* + A^{*0}\mathbf{A})\nonumber\\
  &+& \left. 2(q\cdot A)(q\cdot A^*)q^0 \mathbf{q} \;\rule{0mm}{4.5mm}\right] \ ,
\end{eqnarray}
where $A$ and $A^*$ represent the incoming and outgoing photon fields, respectively.

\begin{table}[!htb]
\begin{center}
\begin{tabular}
{|c|c|c|}\hline
\label{NRQED_photon_Table}
 {\rm operator} & {\it local} & {\it non-local} \\
 \hline\hline
 $\mathbf{x} \times ( \mathbf{E} \times \mathbf{B} )$ & $i\left[(\mathbf{A} \times \mathbf{A}^*)q^0 + (\mathbf{A}\times\mathbf{q})A^{*0} \right.$ & $-2(A \cdot A^*)q^0\mathbf{q}$\\
  & $ \left.+ 2 (\mathbf{q}\times\mathbf{A}^*)A^0 \right]$ & $ + (q\cdot A)q^0 \mathbf{A}^* + (q\cdot A^*)q^0 \mathbf{A}$ \\
 \hline
$ (\p^0 F^{\rho i}) F_\rho^{\;j} $ & $ -2i\left[ q^0 q^2\ma{A}\times\ma{A}^*  \right.$ & $-$ \\
  & $\left. -(q\cdot A^*) q^0 \ma{A}\times\ma{q} - (q\cdot A)q^0 \ma{q}\times\ma{A}^*  \right]$ & \\
 \hline
$ (\p^i F^{\rho0}) F_\rho^{\;j} $ & $-i\left[(q^2 A^0 -q\cdot A q^0) \ma{q} \times \ma{A}^*\right.$ & $-$ \\
 & $\left. + (q^2 A^{*0} -q\cdot A^* q^0) \ma{A} \times \ma{q}\right]$ & \\
 \hline
$ (\square F^{0\rho}) F_\rho^{\;j} x^i $ & $ -i\left[ 4 q^2 A^0 \ma{q}\times \ma{A}^* - q^2 A^{*0} \ma{A}\times\ma{q}\right. $ & $  (q^2)^2 (A^{*0}\ma{A} + A^0\ma{A}^*)  $\\
 & $\left. +q^0 q^2 \ma{A}\times\ma{A}^* - 2 q\cdot A q^0 \ma{q}\times \ma{A}^*\right]$ & $ + 2q^0 q^2 (A\cdot A^*)\ma{q} -q^2(q\cdot A)(q^0 \ma{A}^* +A^{*0}\ma{q}) $ \\
  & & $-q^2(q\cdot A^*)(q^0 \ma{A} +A^{0}\ma{q})$\\
 \hline
$ (\p^0 F^{\rho\sigma}) (\p^j F_{\rho\sigma}) x^i $ & $4i q^0(q\cdot A)\ma{q}\times\ma{A}^* $ & $4 q^0 q^2(A\cdot A^*)\ma{q} - 4q^0(q\cdot A)(q\cdot A^*)\ma{q} $ \\
 \hline
\end{tabular}
\caption{{\it Local} and {\it non-local} contributions to off-shell matrix element $\la\gamma|\m{O}^{\rm eff}|\gamma\ra_{\rm Tree}$.}
\end{center}
\end{table}

In NRQED, since the energy fluctuation cannot excite the virtual electron-position pair, there are no loop corrections to the matrix elements of the effective operators.
All the loops in QED will be encoded in the corresponding Wilson coefficients.
All the tree-level matrix elements of the effective operators in Eqs.~(\ref{spin}), (\ref{orbit}) and (\ref{gamma}) are summarized Table II.
Matching the matrix elements between QED and NRQED, it is easy to find the follow combinations of effective operators
\begin{eqnarray}
 ( \mathbf{S}_q^3)^{(1)} &\to& \frac{\alpha_{\rm _{EM}}}{2\pi} \frac{1}{4m^2} \epsilon^{ij3} \left( - \frac{2}{3} \right) \left[ (\p^0 F^{\rho i}) F_\rho^{\;j} - 2 (\p^j F^{\rho0}) F_{\rho}^{\;j} \right]  \ , \nonumber \\
 ( \mathbf{L}_q^3)^{(1)} &\to& \frac{\alpha_{\rm _{EM}}}{2\pi} \frac{1}{4m^2} \epsilon^{ij3} \left[ - \frac{8}{15} (\square F^{0\rho}) F_\rho^{\; j} x^i - \frac{4}{15} (\p^0 F^{\rho\sigma}) (\p^j F_{\rho\sigma}) x^i + \frac{2}{15} (\p^0 F^{\rho i}) F_\rho^{\;j} - \frac{4}{3} (\p^iF^{\rho0}) F_\rho^{\;j} \right] \ . \nonumber \\
\end{eqnarray}
A little algebra yields the operator relations
\begin{eqnarray}
(\partial^0 F^{\rho i})F_\rho^{\;j} &\to& \dot{\mathbf{E}}\times\mathbf{E} - \dot{\mathbf{B}}\times\mathbf{B} \ , \nonumber \\
(\p^iF^{k0}) F_k^{\;j} &\to& (\bg{\nabla}\cdot\ma{E}) \ma{B} - (\ma{B}\cdot\bg{\nabla})\ma{E} \ , \nonumber \\
\square F^{0\rho}F_\rho^{\;j} x^i &\to& - \mathbf{x} \times ( \square \mathbf{E} \times \mathbf{B} ) \ , \nonumber \\
\partial^0 F^{\rho\sigma}\partial^j F_{\rho\sigma} x^i &\to& 2 \left[ \dot{\mathbf{E}}^a (\mathbf{x}\times \bg{\nabla}) \mathbf{E}^a - \dot{\mathbf{B}}^a(\mathbf{x}\times \bg{\nabla}) \mathbf{B}^a \right] \ .
\end{eqnarray}
Together with Eqs.~(\ref{spin}) and (\ref{orbit}), we extract the Wilson coefficients in the photon sector in the $\overline{\rm MS}$ scheme
\begin{eqnarray}
\label{WilsonGauge}
&&a_{\gamma_1}=-\frac{\alpha_{\rm _{EM}}}{12\pi}, \ \ \  a_{\gamma_2}=\frac{\alpha_{\rm _{EM}}}{6\pi} \ ,\nonumber\\
&&  d_\gamma=\frac{\alpha_{\rm _{EM}}}{3\pi}\ln\frac{\mu^2}{m^2}, \ \ \  d_{\gamma_1} = \frac{\alpha_{\rm _{EM}}}{60\pi}, \ \ \ d_{\gamma_2}= -\frac{\alpha_{\rm _{EM}}}{6\pi}, \ \ \  d_{\gamma_3}= - \frac{\alpha_{\rm _{EM}}}{15\pi}, \ \ \  d_{\gamma_4}= - \frac{\alpha_{\rm _{EM}}}{15\pi} \ . \nonumber\\
&&  f_{\gamma}=1 \ .
\end{eqnarray}

\section{Distinguishing QED and NRQED UV Cutoffs}

In the previous sections, we have obtained the non-relativistic reduction
of the angular momentum operators with the main results in Eqs.~(\ref{spin})-(\ref{gamma}) and
Eqs.~(\ref{WilsonBilinear}), (\ref{WilsonGauge}). We have used the dimensional regulation for both UV and IR divergences
in QED and NRQED, and the Wilson coefficients depend on a single energy scale $\mu$. However, there is no
need that the NRQED must use the same UV regularization. In fact, in many bound state calculations,
it is better to use three-momentum cut-off for the UV divergences and the photon mass for IR divergences.
In this section, we will try to redo the calculations in this new regularization scheme.  Before that, let's
consider a problem that arises from DR for all divergences.

We expect the UV behaviors of QED operators are completely included in the Wilson
coefficients in NRQED. In another word, the matrix elements in NRQED must
satisfy the same renormalization group (RG) evolution equations
\begin{equation}
\label{RG_Equation}
\frac{d}{d t}{\ma{J}_q^{\rm eff} \choose \ma{J}_\gamma^{\rm eff}} = \frac{\alpha_{\rm _{EM}}}{2\pi} {-\frac{4}{3}\;\;\;\;\,\frac{1}{3} + \frac{1}{3} \choose \;\;\,\frac{4}{3}\;\;-\frac{1}{3} + \frac{1}{3}} {\ma{J}_q^{\rm eff} \choose \ma{J}_\gamma^{\rm eff}} \ ,
\end{equation}
where $t = \ln \mu^2/m^2$ and $\ma{J}_q^{\rm eff} \equiv \ma{S}_q^{\rm eff} + \ma{L}_q^{\rm eff}$ represents the total angular momentum carried by the electron. The additional $\frac{1}{3}$ in the evolution equation stems from the scale dependence of the redefined photon fields in the effective theory, the $d_1$ term.

Our effective operators in Eqs.~(\ref{spin})-(\ref{gamma}) with
Wilson coefficients listed in Eqs.~(\ref{WilsonBilinear}) and
(\ref{WilsonGauge}), actually, fail to satisfy the desired
evolutions. The reason for this, as already discussed in subsection
III, is due to DR. In the QED
calculations, we have not distinguish the $\mu$ dependence from UV
or IR divergences. The IR divergences should be captured by the
effective theory and has nothing to do with the RG flow. On the
other hand, when matching to NRQED, we also have not distinguished the
UV cutoff dependence from that in QED. By introducing the
three-momentum cutoff $\Lambda$ in NRQED calculations, we can restore the correct RG evolutions.
In addition, We let the photon have a small mass $\lambda$ in
both QED and NRQED to regulate the infrared. According to
Ref.~\cite{Kinoshita:1995mt} the Wilson coefficients in NRQED
Lagrangian with three-momentum cutoff $\Lambda$ as UV regulator are
\begin{eqnarray}
\label{cutoff_Lag}
  &&c_2 = c_4 = 1,\;\; c_F = 1 + \frac{\alpha_{\rm _{EM}}}{2\pi},\;\; c_D = 1 - \frac{8\alpha_{\rm _{EM}}}{3\pi} \left(\ln\frac{2\Lambda}{m} - \frac{5}{6}\right),\;\; c_S = 1 + \frac{\alpha_{\rm _{EM}}}{\pi} \ ,\nonumber\\
  &&d_1 = 1 + \frac{\alpha_{\rm _{EM}}}{3\pi}\ln\frac{\mu^2}{m^2},\;\; d_2 = \frac{\alpha_{\rm _{EM}}}{60\pi} \ .
\end{eqnarray}
The separation of two scales $\mu$ and $\Lambda$ is obvious.

\begin{figure}
\begin{center}
\includegraphics[width=3cm]{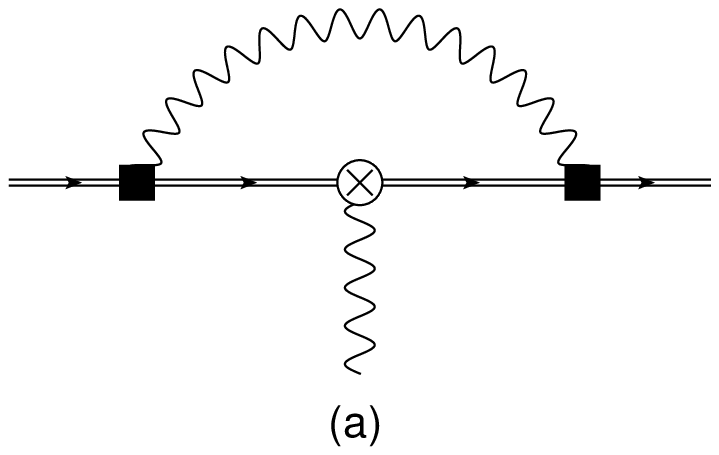}
\hspace{2mm}
\includegraphics[width=3cm]{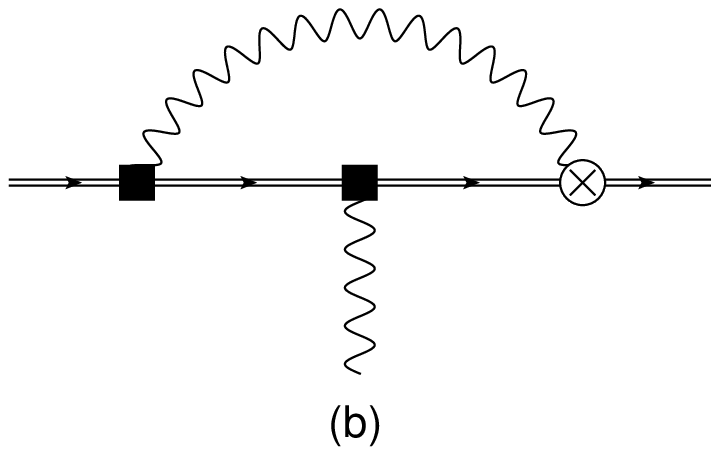}
\hspace{2mm}
\includegraphics[width=3cm]{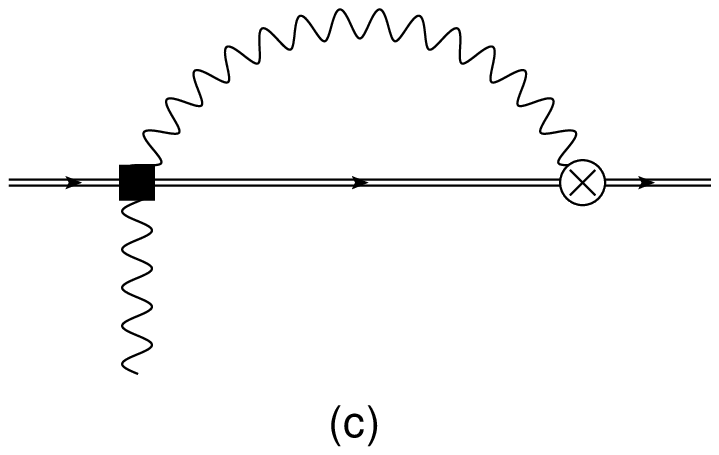}
\hspace{2mm}
\includegraphics[width=3cm]{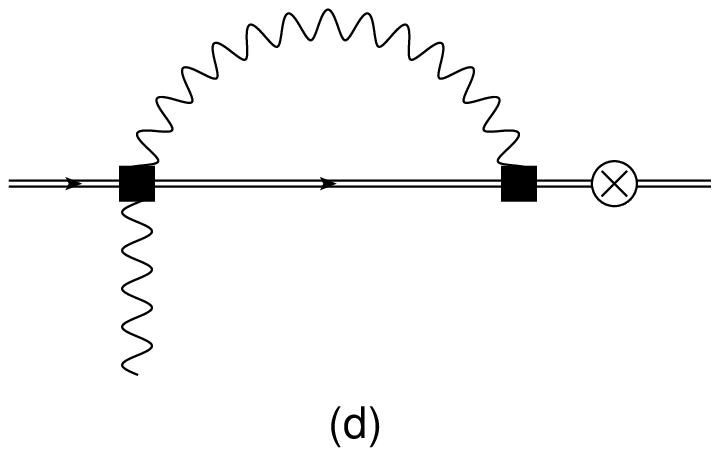}\\
\vspace{2mm}
\includegraphics[width=3cm]{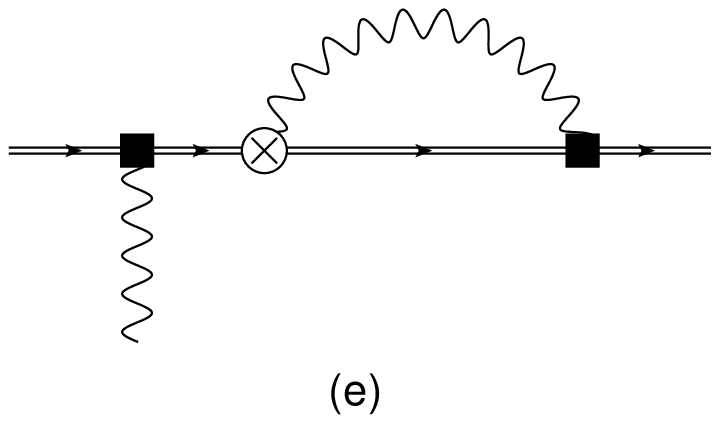}
\hspace{2mm}
\includegraphics[width=3cm]{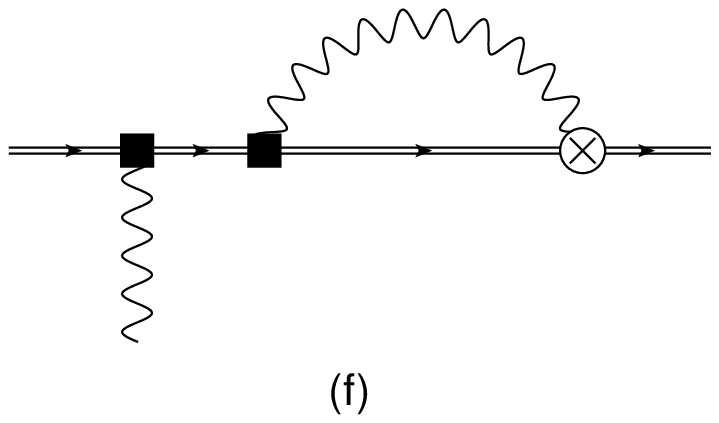}
\hspace{2mm}
\includegraphics[width=3cm]{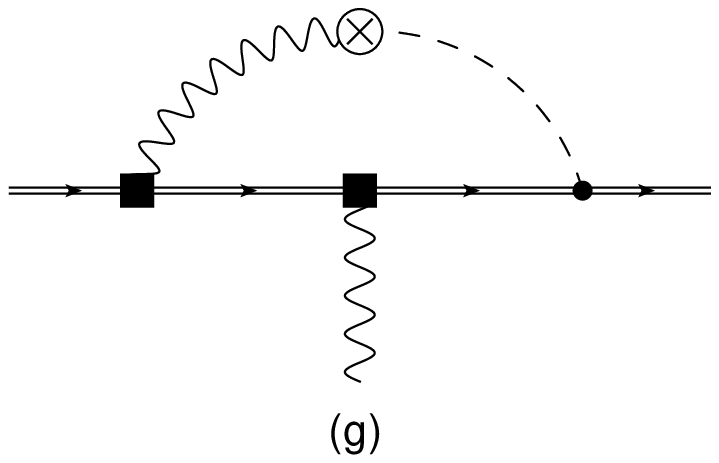}
\hspace{2mm}
\includegraphics[width=3cm]{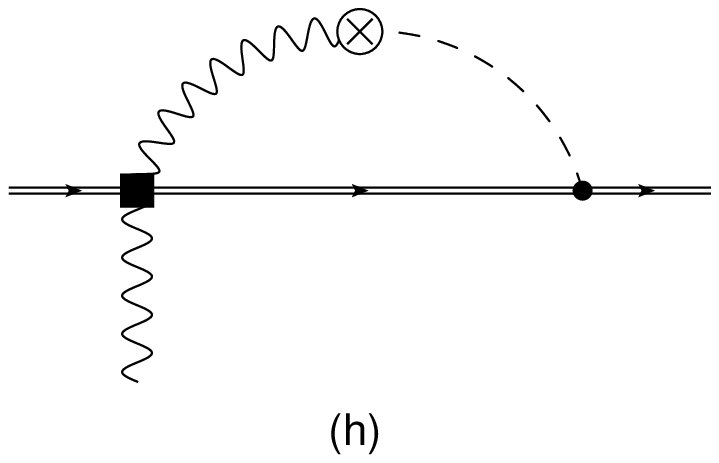}\\
\vspace{2mm}
\includegraphics[width=3cm]{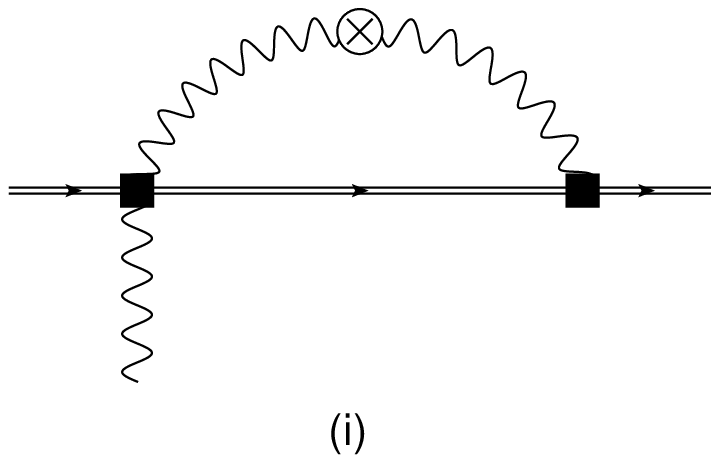}
\hspace{2mm}
\includegraphics[width=3cm]{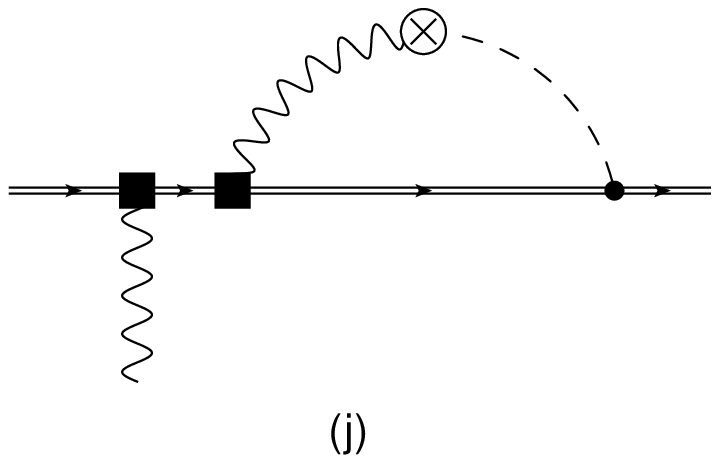}
\hspace{3mm}
\includegraphics[width=3cm]{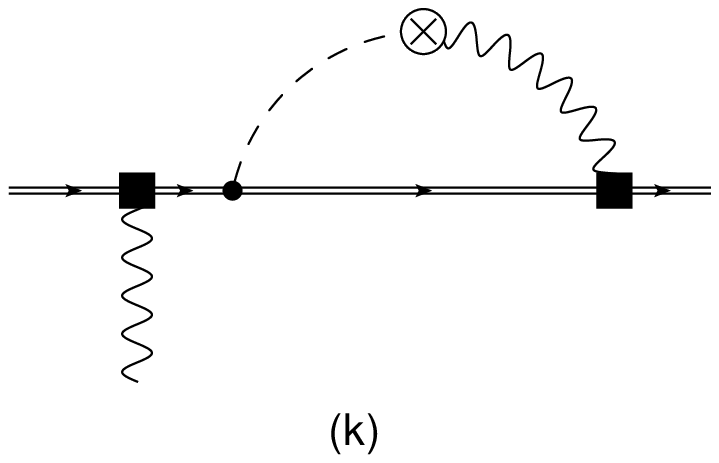}
\caption{One-loop corrections to $\la e|\ma{L}_q^{\rm eff}|e\ma{A}\ra$, (a)-(f), and $\la e|\ma{J}_\gamma^{\rm eff}|e\ma{A}\ra$, (g)-(k), in NRQED. In the diagram the filled box represents the vertex interaction from term $-\frac{\ma{D}^2}{2m}$ in the NRQED Lagrangian. In the diagram the filled box represents the vertex interaction from term $-\frac{\ma{D}^2}{2m}$ in the NRQED Lagrangian and the dot represents the $D^0$ vertex. The dashed line is for the Coulomb photon propagator while the wavy line is for the transverse photon propagator.  Wavefunction renormalization diagrams, mass counterterms and the mirror diagrams are not shown explicitly.  }
\label{NR_OneLoop}
\end{center}
\end{figure}

The full theory amplitude with the photon mass regulator can be translated by making the replacement in Eqs.~(\ref{S_spin})-(\ref{g_orbital})
\begin{equation}
\left(\frac{1}{\e_{\rm IR}} + \ln\frac{\mu^2}{m^2}\right) \rightarrow 2\ln\frac{\lambda}{m} \ .
\end{equation}
In the effective theory, the loop diagrams are no longer vanishing if we use $\Lambda$ and $\lambda$ to regulate UV and IR divergences.
We choose to work in the Coulomb gauge to do the calculation since the full result is gauge-invariant.
All the diagrams under consideration in NRQED are listed in Fig.~\ref{NR_OneLoop}.
The non-relativistic electron propagator picks up only the pole at $p^0=\sqrt{\ma{p}^2+m^2}$.

\begin{equation}
S_F^{\rm eff}(p+k) = \frac{1}{E_p+k^0-\frac{(\ma{p}+\ma{k})^2}{2m}+ i\epsilon} = \frac{1}{k^0+i\epsilon} + \frac{2\ma{p}\cdot\ma{k}+\ma{k}^2}{2m(k^0+i\epsilon)^2} + \cdots .
\end{equation}
This can be understood as a multipole expansion in inverse powers of $m$.

The one-loop effective theory matrix elements now read:
\begin{eqnarray}
\la \ma{L}_q^{\rm eff} \ra^{(1)}_{\rm local} & = & \frac{\alpha_{\rm _{EM}}}{2\pi} \frac{-4ie}{m^2} d_R^{(0)} \left(\ln\frac{2\Lambda}{\lambda} - \frac{5}{6} \right) u_h^\dag\ma{q}\times\ma{A}u_h \ , \nonumber\\
\la \ma{J}_\gamma^{\rm eff} \ra^{(1)}_{\rm local} & = & \frac{\alpha_{\rm _{EM}}}{2\pi} \frac{4ie}{m^2} f_\gamma^{(0)} \left(\ln\frac{2\Lambda}{\lambda} - \frac{5}{6} \right) u_h^\dag\ma{q}\times\ma{A}u_h \ , \nonumber\\
\la \ma{L}_q^{\rm eff} \ra^{(1)}_{\rm non-local} & = & \frac{\alpha_{\rm _{EM}}}{2\pi} \frac{2}{3m^2} d_R^{(0)} \left(\ln\frac{2\Lambda}{\lambda} - \frac{5}{6} \right)\ma{x}\times u_h^\dag |\ma{q}|^2\ma{A} u_h \ .
\end{eqnarray}
The one-loop amplitude $\la \ma{S}_q^{\rm eff} \ra_{\rm local}$ and $\la \ma{J}_\gamma^{\rm eff} \ra_{\rm non-local}$ as well as the diagrams involving the scalar potential $A^0$ vanish at higher order $\m{O}(m^{-2})$ in NRQED.

Now the matching condition Eq.~(\ref{MatchingCondition}) should be rewritten as
\begin{equation}
\label{MatchingConditionNew}
\la\mathbf{J}(\mu)\ra_{\rm QED} = \la\mathbf{J}^{\rm eff}(\mu,\Lambda)\ra_{\rm NRQED} \ .
\end{equation}
The sum rules of the Wilson coefficients in Eq.~(\ref{sum_rule}) do not change. However, all the Wilson coefficients in the effective Lagrangian are defined in the presence of new regulators, similar to Eq.~(\ref{cutoff_Lag}).

With the new matching conditions, we find yet another set of Wilson coefficients depending on both three-momentum cutoff $\Lambda$ in NRQED and the QED cutoff $\mu$.  We only list the ones different from those in Eq.~(\ref{WilsonBilinear}), and denote them with an asterisk.
\begin{eqnarray}
&&d_D^* = 1 + \frac{\alpha_{\rm _{EM}}}{2\pi} \left(-\frac{4}{3} \ln\frac{\mu^2}{m^2} + \frac{16}{3} \ln\frac{m}{2\Lambda} + \frac{37}{9}\right) \ ,\nonumber\\
&&d_B^* = -1 + \frac{\alpha_{\rm _{EM}}}{2\pi} \left( -16\ln\frac{m}{2\Lambda} + \frac{23}{9}\right)\ ,\nonumber \\
&&f_B^* = \frac{\alpha_{\rm _{EM}}}{2\pi} \left( 16\ln\frac{m}{2\Lambda} - \frac{86}{9}\right) \ .
\end{eqnarray}
With the new coefficients, both $\ma{J}_q^{\rm eff}$ and $\ma{J}_\gamma^{\rm eff}$ now satisfy the RG equations, Eq.~(\ref{RG_Equation}). They now agree with those in Eq.~(8) in Ref.~\cite{Chen:2009rw}.

\section{Conclusions}

In this paper, we have established systematically the relativistic
angular momentum components in the framework of NRQED up
to the order $\alpha_{\rm _{EM}}/m^2$. Such effective operators can be
used in the computation of higher-order contributions to
spin/orbital angular momentum in non-relativistic bound states, such
as the hydrogen atom. Further extensions to the effective operators in
NRQCD can be readily obtained following the same procedure, and
will be useful for studying the strong-interaction bound states
such as the heavy quarkonium systems.

We would like to thank Peng Sun and Yang Xu for discussions and comments on the manuscripit.
This work was partially supported by the U. S. Department of Energy via grant DE-FG02-93ER-40762. 
YZ acknowledges the support from the TQHN group at University of Maryland and the Center for High-Energy 
Physics at Peking University where part of the work was done.

\appendix
\section{Generalized N\"{o}ether Currents with Higher Derivatives}

If the effective theory Lagrangian contains higher derivatives terms the conventional equation of motion and N\"{o}ether currents, derived from $\m{L}=\m{L}(\phi,\p^\mu\phi)$, have to be modified.
In this appendix, following the procedures in Ref.~\cite{book}, we have generalized the formulas to a Lagrangian with second-order derivative $\m{L}=\m{L}(\phi,\p^\mu\phi,\p^\mu\p^\nu\phi)$, which will be relevant to deriving the angular momentum in NRQED, Eq.~(\ref{NRQED_J}). We list the results here.

We use the following notations:
\begin{equation}
  \m{L}_\alpha \equiv \frac{\p\m{L}}{\p \phi^\alpha},\;\; [\m{L}_\alpha]^\mu \equiv \frac{\p\m{L}}{\p(\p_\mu \phi^\alpha)}, \;\; [\m{L}_\alpha]^{\mu\nu} \equiv \frac{\p\m{L}}{\p(\p_\mu \p_\nu \phi^\alpha)} \ .
\end{equation}
The generalized equation of motion is
\begin{equation}
  \m{L}_\alpha - \p_\mu[\m{L}_\alpha]^\mu + \frac{1}{2}\p_\mu\p_\nu [\m{L}_\alpha]^{\mu\nu} = 0 \ .
\end{equation}
The generalized, Belinfante-improved, symmetric energy-momentum tensor $T^{\mu\nu}$ has the form
\begin{equation}
  T^{\mu\nu}  =  -\m{L}g^{\mu\nu} + [\m{L_\alpha}]^\mu \p^\nu \phi^\alpha - \frac{1}{2}\p_\eta[\m{L_\alpha}]^{\mu\eta} \p^\nu \phi^\alpha +\frac{1}{2}[\m{L_\alpha}]^{\mu\eta}\p_\eta\p^\nu \phi^\alpha - \p_\sigma f^{\mu\sigma\nu} \ ,
\end{equation}
where $f^{\mu\sigma\nu}$ is defined as
\begin{eqnarray}
  f^{\mu\sigma\nu} & \equiv & \frac{1}{2}\left[([\m{L_\alpha}]^\mu \phi_\beta - \frac{1}{2}\p_\eta[\m{L_\alpha}]^{\mu\eta} \phi_\beta + \frac{1}{2}[\m{L_\alpha}]^{\mu\eta} \p_\eta\phi_\beta)(S^{\alpha\beta})^{\sigma\nu} - (\mu\leftrightarrow\sigma) + (\mu\leftrightarrow\nu) \right]\nonumber\\
  & & - \frac{1}{2}[\m{L_\alpha}]^{\mu\nu} \p^\sigma\phi^\alpha + \frac{1}{2}[\m{L_\alpha}]^{\sigma\nu} \p^\mu\phi^\alpha \ ,
\end{eqnarray}
where $(S^{\alpha\beta})^{\sigma\nu}$ is the generator of the Lorentz group.
Finally, the generalized angular momentum tensor can be obtained through
\begin{eqnarray}
  \m{M}^{\mu\nu\lambda} & = & T^{\mu\lambda}x^\nu - T^{\mu\nu}x^\lambda, \;\;\;  \ma{J}=\epsilon^{ij3} \int d^3x \m{M}^{0ij} \ .
\end{eqnarray}

As an application to NRQED, for the higher derivative operator $\frac{d_2}{m^2} F_{\mu\nu}\square F^{\mu\nu}$ in the Lagrangian, we obtain the corresponding
angular momentum operators
\begin{eqnarray}
&&\left(-\frac{2d_2}{m^2}\right)\epsilon^{ij3}
\left[2\square F^{0\rho}F_\rho^{\;j} x^i + \partial^0 F^{\rho\sigma}\partial^j F_{\rho\sigma} x^i + 2(\partial^0 F^{\rho i})F_\rho^{\;j} \rule{0cm}{4mm}\right] \\
&&= \left(-\frac{4d_2}{m^2}\right) \left[ - \mathbf{x} \times ( \square \mathbf{E} \times \mathbf{B} ) + \dot{\mathbf{E}}^a (\mathbf{x}\times \bg{\nabla}) \mathbf{E}^a - \dot{\mathbf{B}}^a(\mathbf{x}\times \bg{\nabla}) \mathbf{B}^a + \dot{\mathbf{E}}\times\mathbf{E} - \dot{\mathbf{B}}\times\mathbf{B}\right] \nonumber \ ,
\end{eqnarray}
which corresponds to the terms proportional to $d_2$ in Eq.~(\ref{NRQED_J}).

The other terms in $\mathcal{L}_{\rm NRQED}$ contains at most one derivative, so the remaining parts of angular momentum effective operators can be readily obtained.

\end{document}